# Unveiling the Miniband Structure of Graphene Moiré Superlattices via Gate-dependent Terahertz Photocurrent Spectroscopy

Juan A. Delgado-Notario[1*], Stephen R. Power[2], Wojciech Knap[3,4], Manuel Pino[5], Jin Luo Cheng[6], Daniel Vaquero[7], Takashi Taniguchi[8], Kenji Watanabe[8], Jesús E. Velázquez-Pérez[1], Yahya Moubarak Meziani[1], Pablo Alonso-González[9], José M. Caridad[1,10*]

[1]Departamento de Física Aplicada, Universidad de Salamanca, 37008 Salamanca, Spain

[2]School of Physical Sciences, Dublin City University, Glasnevin, Dublin 9, Ireland

[3]CENTERA Labs, Institute of High Pressure Physics, Polish Academy of Sciences, Warsaw 01-142, Poland.

[4]Centre for Advanced Materials and Technologies CEZAMAT, Warsaw University of Technology, Warsaw 02-822, Poland.

[5]Departamento de Física Fundamental (IUFFyM) y GIR Nanotecnología, Universidad de Salamanca, 37008 Salamanca, Spain

[6]GPL Photonics Laboratory, State Key Laboratory of Luminescence and Applications, Changchun Institute of Optics, Fine Mechanics and Physics, Chinese Academy of Sciences, Changchun, Jilin 130033, People's Republic of China and University of Chinese Academy of Sciences, Beijing 100039, China

[7]Zernike Institute for Advanced Materials, University of Groningen, 9747 AG Groningen, The Netherlands

[8]Research Center for Electronic and Optical Materials, National Institute for Materials Science, Tsukuba 305-0044, Japan

[9]Department of Physics, University of Oviedo, Oviedo 33006, Spain

[10]Unidad de Excelencia en Luz y Materia Estructurada (LUMES), Universidad de Salamanca, Salamanca 37008, Spain

\* *emails*: juanandn@usal.es, jose.caridad@usal.es

ABSTRACT

Moiré superlattices formed at the interface between stacked two-dimensional atomic crystals offer limitless opportunities to design materials with widely tunable properties and engineer intriguing quantum phases of matter. However, despite progress, precise probing of the electronic states and tantalizingly complex band textures of these systems remain challenging. Here, we present gate-dependent terahertz photocurrent spectroscopy as a robust technique to detect, explore and quantify intricate electronic properties in graphene moiré superlattices. Specifically, using terahertz light at different frequencies, we demonstrate distinct photocurrent regimes evidencing the presence of avoided band crossings and tiny (~1-20 meV) inversion-breaking global and local energy gaps in the miniband structure of minimally twisted graphene and hexagonal boron nitride heterostructures, key information that is inaccessible by conventional electrical or optical techniques. In the off-resonance regime, when the radiation energy is smaller than the gap values, enhanced zero-bias responsivities arise in the system due to the lower Fermi velocities and specific valley degeneracies of the charge carriers subjected to moiré superlattice potentials. In stark contrast, above-gap excitations give rise to bulk photocurrents– intriguing optoelectronic responses related to the geometric Berry phase of the constituting electronic minibands. Besides their fundamental importance, these results place moiré superlattices as promising material platforms for advanced, sensitive and low-noise terahertz detection applications.

INTRODUCTION

In van der Waals (vdW) heterostructures, lattice mismatch and rotation between adjacent crystals can lead to the formation of a moiré superlattice with a periodicity larger than the atomic scale. Such periodic potentials induce notable changes in the electronic, optical and mechanical properties of different two-dimensional (2D) crystals[1-3], and may prompt the appearance of a number of exotic fundamental phenomena. In the case of graphene-based moiré superlattices, some of these effects include, for instance, superconductivity[4], magnetism[5] or correlated insulator phases[6]. In addition, moiré structures have recently been shown to be an interesting material platform for the realization of different electronic components such as single-electron transistors[7], superconducting quantum interference devices[8] or efficient visible/infrared photodetectors[9,10].

According to theoretical predictions[11-13], graphene moiré superlattices are also expected to be unique systems to design state-of-the-art, compact optoelectronic devices working at terahertz (THz) frequencies[14,15]. Moreover, measurements in this frequency range also have the potential to act as a diagnostic tool, providing valuable and elusive information about the unique and convoluted electronic structure of moiré superlattice systems[13,16-18] which ultimately depends on the subtle atomic registry and interactions existing between the layered materials.

A relevant example is the absence or presence (and possible quantification) of avoided band crossings and tiny (<10 meV) energy gaps in the conduction band states of elementary moiré superlattices made of graphene and hexagonal boron nitride, intriguing features which have remained under debate up to now[2,19,20]. In this sense, a key open question is whether such energy gaps do exist but have not been experimentally observed due to the insufficient sensitivity and resolution of the applied experimental methods and diagnostic tools so far.

In the present work, we combine THz photocurrent spectroscopy (excitation frequencies between 0.075 and 4.7 THz) with the continuous tuning of the Fermi level via gate voltage, to explore and probe the band structure singularities and intricate electronic states of graphene moiré superlattices. Our samples were fabricated by aligning the crystal lattices of graphene and hexagonal boron nitride (hBN) at rotation angles θ < 2° (**Figure 1a**). Interestingly, we show that the gate-tuneable THz photocurrent collected in these devices can originate from intraband or interband transitions, depending on the excitation frequency, and such responses are exquisitely sensitive to very fine details of the electronic structure of graphene in the presence of moiré potentials. Specifically, THz photocurrent measurements at multiple frequencies provide important information about the miniband structure appearing in the vicinity of the $\bar{K}/\bar{K}'$ and $\Gamma$ points of the superlattice Brillouin zone, featuring a new generation of Dirac fermions close to the so-called secondary or superlattice Dirac points, sDPs[1-3] (**Figure 1b**) as well as other important aspects such as the reduced Fermi velocity of charge carriers near sDPs with respect to the main Dirac point, DP[21], or the presence and size of energy gaps at the satellite Dirac point of the valence band[2] ($\Delta_h$). More intriguingly, we also demonstrate how gate-tunable THz photoresponse measurements of graphene moiré superlattices at multiple frequencies are able to detect the presence of tiny local energy gaps (sizes of few meV) at the satellite Dirac point of the conduction band ($\Delta_e$). These are subtle features "hidden" to conventional probing techniques such as quantum transport measurements[2,20] or angle-resolved photoemission spectroscopy (ARPES)[22], experimental methods which can only probe overall gaps (i.e. bandgaps) existing in these quantum materials or are limited to an energy resolution of ~30 meV, respectively. In addition, we unveil that the recorded photoresponses are furthermore sensitive to quantum geometric textures of the gapped states occurring at the miniband satellite Dirac points.

As such, the gate- and frequency-dependent THz photoresponse of graphene moiré materials is demonstrated to be a complete and sensitive technique to accurately probe and disclose relevant and exclusive information about the intricate electronic structure of graphene's Dirac electrons subjected to superlattice potentials. This outstanding ability is of particular significance given the fact that current fabrication methods suffer from distortions (due to random strain or the control of the twist angle[23]) that impact the electronic properties of graphene moiré superlattices. From a more technological point of view, samples where the crystal lattices of graphene and hBN are close to perfect alignment ($\theta = 0°$) show enhanced zero drain-bias responsivity values close to the sDPs with respect to the main DP, and thus this work also paves the way towards engineering moiré superlattice based THz detectors with high-speed, low-noise and extreme sensitivity.

## RESULTS AND DISCUSSION

**Fabrication of moiré THz devices**

Following published techniques for the exfoliation and van der Waals assembly of 2D materials[24,25], we assemble (***Methods and Supporting Information Note 1***) several graphene heterostructures, where both the graphene (monolayer, MLG, or bilayer, BLG) and hBN lattices are aligned by a rotation angle $\theta < 2°$ (see Figure 1a). We anticipate that the use of monolayer or bilayer graphene crystals in our samples is equivalent for the present study (see electrical and photocurrent measurements described below).

We then fabricate five different devices and measure their zero drain-bias photocurrent at THz frequencies (see **Figure 1c** and ***Methods***). The top-left inset of **Figure 2a** shows one of these devices (here named device A), consisting of a conventional short-channel (SC), dual-gated architecture[26-28]. Besides, we measure four additional devices shaped in a similar SC architecture (device B) or other different and well-known geometries (e.g. multicross bars[14] (MC), device C, or interdigitated dual-gated transistors[29] (IDGT), devices D and E, see details in **Supporting Information Note 2**) to ensure that our conclusions are generic to the effect of moiré potentials in graphene and not dependent on a specific device architecture.

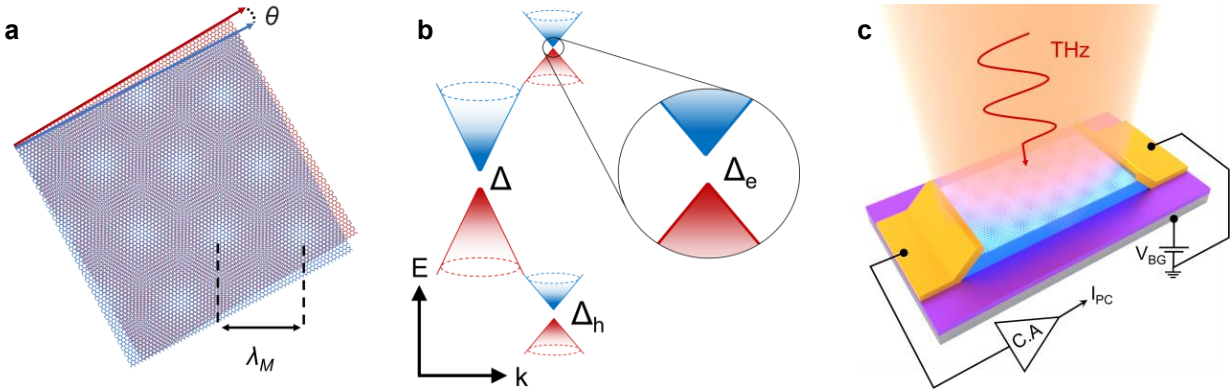

*Figure 1. Moiré THz devices. a,* Illustration of a moiré pattern with long wavelength $\lambda_M$, generated by aligning graphene (red) and hBN (blue) honeycomb lattices by an angle $\theta$ below 2 degrees. *b,* Schematic of the bandstructure of the graphene/hBN moiré superlattices featuring a new generation of Dirac fermions (so-called secondary or superlattice Dirac points, sDPs), the appearance of gaps at the main ($\Delta$) and superlattice Dirac points ($\Delta_e$ and $\Delta_h$) and the different Fermi velocity of charge carriers close to each of the Dirac points (i.e. different slope of the Dirac cones)[21]. *c,* Schematic of the zero drain-bias photocurrent ($I_{PC}$) measurements taken at different THz frequencies.

**Electrical characterization of the devices**

We first measure and assess the transport characteristics of the fabricated devices (measurement details shown in *Methods*). All these systems are of high quality, with carrier mobilities well exceeding 150000 cm$^2$V$^{-1}$s$^{-1}$ at low temperatures $T$ (see ***Supporting Information Note 3***). **Figure 2** depicts the measured channel resistance, r$_{ch}$, as a function of the carrier density, *n*, for the SC devices of the study, devices A (**Figure 2a**) and B (**Figure 2b**) at 10 K. The corresponding data for devices C, D and E are shown in ***Supporting Information Note 2***. In agreement with previous transport measurements of MLG and BLG coupled to a moiré superlattice[2,3], r$_{ch}$ in all these devices exhibits three resistance peaks at carrier densities 0 cm$^{-2}$ and ±|$n_{SP}$| (the latter positioned around ~ ±8×10$^{12}$ cm$^{-2}$ and ~ ±4×10$^{12}$ cm$^{-2}$ for devices A and B, respectively). Such features are the result of a depression in the density of states occurring in the electronic band-structure of these systems at the DP and the two sDPs (in both conduction and valence bands), respectively.

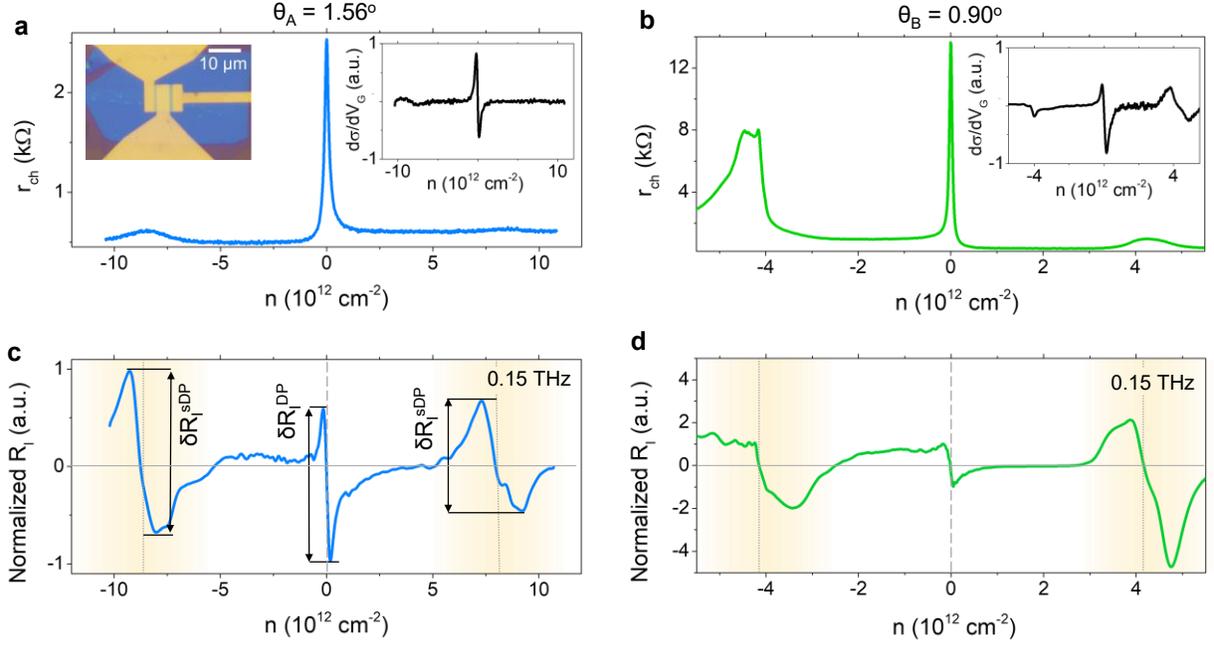

*Figure 2. Electrical and optoelectronic characteristics of graphene moiré devices. a, Measured channel resistance, $r_{ch}$, as a function of the carrier density, n, for device A at 10K. Top left inset shows an optical photograph of a SC device. b, Measured channel resistance, $r_{ch}$, as a function of the carrier density, n, for device B at 10K. Top right insets in panels a and b show the variation of the channel conductivity σ with respect to the carrier density, dσ/dn, as a function of n. c, Measured photocurrent responsivity $R_I$ as a function of the carrier density, n, for device A at 10K under the excitation of 0.15 THz. d, Measured photocurrent responsivity $R_I$ as a function of the carrier density, n, for device B at 10K under the excitation of 0.15 THz. Vertical lines mark the position of the main (dashed) and satellite (continuous) Dirac points.*

By using the charge density at which these satellite resistance peaks are observed $n_{SP}$, one can estimate both the moiré wavelength, $\lambda_M$, and the relative rotation angle, θ, between the graphene and hBN lattices in the devices[1-3]. In particular, $\lambda_M^{(A)}$ = 7.6 nm, $\theta_A$ = 1.56°, and $\lambda_M^{(B)}$ = 10.4 nm, $\theta_B$ = 0.9° for devices A and B, respectively. The details of these calculations as well as the values of $\lambda_M$ and $\theta$ extracted for all devices A-E are shown in *Supporting Information Information Note 4*. In general, $\theta$ lies between 0.4° and 1.6° in all studied devices. We note that many of our devices show long wavelength values which are not much smaller than the expected maximum around 14 nm[1,2], which implies a nearly perfect rotational alignment of the graphene and hBN crystals with $\theta < 1°$ and may result in a commensurate state of graphene on hBN (local enlargement of the lattice constant of graphene to match the one of hBN).

**THz photodetection in graphene moiré superlattice devices**

Figures **2c** and **2d** show the current responsivity, $R_I = I_{PC}/P$, of the fabricated moiré THz detectors A and B, respectively, as a function of *n* measured at 0.15 THz and a temperature of 10K ($R_I$ of devices C, D and E can be found in *Supporting Information Note 5*). Here, $I_{PC}$ is the collected zero drain-bias photocurrent (**Figure 1c**) and *P* is the incident radiation power. For clarity, $R_I$ is normalized with respect to the maximum photocurrent value measured close to the main Dirac point. At a low chemical potential $E_F$, in the vicinity of the main DP (doping level interval $|n| < 2\times10^{12}$ cm$^{-2}$), the photoresponse in the studied devices exhibits a different sign which depends on the type of charge carrier existing in the device channel, electrons or holes. The sign reversal of the photocurrent occurs right at the DP and the photoresponse tends to zero at both ends of the considered interval, i.e. at values $n \sim \pm 2\times10^{12}$ cm$^{-2}$. This qualitative behavior is consistent with the one reported for THz photodetectors made of bare graphene[28-31], and stems from the ambipolar charge transport present in this material, indicating the dominant role of intraband

absorption in the photocurrent $I_{PC}$[28-30]. Interband transitions would result in the generation of electron–hole pairs and the variation of the responsivity with respect to *n* should have yielded a photocurrent peak at the Dirac point[14], instead. In particular, intraband-type photoresponse has a functional dependence on *n* which is proportional to $d\sigma/dn$ (see inset panels of **Figures 2a** and **2b**), with $\sigma$ being the dc conductivity of the material. There are two physical intraband phenomena that may give rise to such type of photoresponse in graphene devices at THz frequencies: plasma wave rectification and photo-thermoelectric effects[30]. The fact that the photoresponse $R_I$ tends to vanish at large *n* (i.e for Fermi levels $E_F$ sufficiently separated from the main Dirac point) indicates that plasma wave rectification is the dominant mechanism in our devices, rather than photo-thermoelectric effects.[26,27,28]

In the following, we examine the measured photoresponse at higher $E_F$, close to the superlattice Dirac points (carrier densities $\pm|n_{SP}|$, see highlighted region in **Figures 2c** and **2d**). In particular, we demonstrate that the lineshape and magnitude of the measured photocurrent responsivity $R_I(n)$ at different frequencies are sensitive to many of the key characteristics of the miniband structure of graphene moiré superlattice systems[32], including: *i)* the presence of superlattice Dirac points, *ii)* the reduced Fermi velocity and distinct valley degeneracy of charge carriers near sDPs with respect to the main Dirac point and even *iii)* the complex band texture occurring at $E_F$ close to the sDPs, including several avoiding band crossings giving rise to a series of global and local energy gaps.

*Presence of superlattice Dirac points.* The overall measured photocurrent response around the position of the satellite Dirac points $\pm|n_{SP}|$ at 0.15 THz exhibits a qualitative trend which is similar to the one reported close to the main Dirac point. In particular, $R_I(n)$ changes sign right at the sDPs both in the hole and the electron bands, $-|n_{SP}|$ and $+|n_{SP}|$ respectively. Such behavior has been already observed[26] and is consistent with the existence of electron-hole pockets located near the $\bar{K}/\bar{K}'$ and $\Gamma$-points of the superlattice Brillouin zone[32] (i.e. superlattice Dirac points) in both valence and conduction bands (**Figure 1b**) and the generation of plasmons in such superlattice minibands[12].

*Reduced Fermi velocity and valley degeneracy of charge carriers near sDPs.* We then examine the measured photoresponses at a more quantitative level. **Figure 2c** shows that, in our device with θ ~1.56˚ (device A), the $R_I$ signal near the two sDPs is similar in magnitude to that measured near the main DP. In contrast, devices with θ <1˚ (devices B-E, see Figure 2d and *Supporting Information Note 5*) present an enhanced $R_I$ around the two sDPs compared to the signal measured near the main DP. We quantify such enhancement by taking into account the photocurrent responsivity variation close to both main ($\delta R_I^{DP}$) and satellite Dirac points ($\delta R_I^{sDP}$), where $\delta R_I^{DP}$ and $\delta R_I^{sDP}$ are the differences in the measured maxima and minima of $R_I$ around carrier densities 0 cm$^{-2}$ and $\pm|n_{SP}|$, respectively (see marks in **Figure 2c**). Overall, enhancement ratios $\delta R_I^{sDP}/\delta R_I^{DP}$ between 1.5 and 5 are observed in devices with θ <1˚ at T=10K (**Figure 2d** and *Supporting Information Note 5*).

Some of these quantitative results are surprising at first glance. In particular, whereas the responsivity ratio $\delta R_I^{sDP}/\delta R_I^{DP} \leq 1$ observed in **Figure 2c** (device A, θ ~1.56˚) can be easily explained by the variation in the channel conductivity with respect to $n$ ($d\sigma/dn$, see inset of **Figure 2a**); ratios $\delta R_I^{sDP}/\delta R_I^{DP} > 1$ observed in **Figure 2d** and *Supporting Information Note 5* (devices B-E, all with θ <1°) cannot be understood from $d\sigma/dn$ (see inset of **Figure 2b** and *Supporting Information Note 6*). The highlighted quantitative difference can be explained by the extraordinary sensitivity of the intraband photocurrent to the electronic Fermi surface of the system under study. In particular, we argue that the $\delta R_I^{sDP}/\delta R_I^{DP} > 1$ measured in devices B-E can be explained by the larger density of states *DOS* present at energies around the sDPs in graphene/hBN heterostructures with small misalignment angles[1] θ<1°. Following a straightforward analysis (see *Supporting Information Note 6*), one can roughly estimate the responsivity enhancement in the zero temperature and zero carrier density limit to be:

$$\frac{\delta R_I^{sDP}}{\delta R_I^{DP}} \approx \frac{\left(g_v^{(sDP)}\right)^{3/2} v_F^{(DP)}}{\left(g_v^{(DP)}\right)^{3/2} v_F^{(sDP)}} \qquad \text{Eq. 1}$$

where $g_v$ and $v_F$ are the valley degeneracy and the Fermi velocity of charge carriers. These two band structure parameters are different near the main and satellite Dirac points;[1,32,33] and such differences can explain the photocurrent enhancement observed near the satellite Dirac points in samples with θ<1°. Indeed, whereas the valley degeneracy near the DP ($g_v^{(DP)}$) is 2, the value near the sDPs ($g_v^{(sDP)}$) is 2 or 6 in common bandstructure reconstructions of graphene moiré superlattices. Moreover, the Fermi velocity of charge carriers at the satellite Dirac points $v_F^{(sDP)}$ in well aligned devices (θ<1°) has been shown to be ~0.50-0.73 times smaller than the Fermi velocity at the Dirac point $v_F^{(DP)}$. From the aforementioned values, one can estimate enhanced responsivity ratios $\delta R_I^{sDP}/\delta R_I^{DP}$ in graphene moiré superlattice devices at vanishing temperature and carrier density to approximately stand between 1 and 10. These calculations are in good agreement with our experiments, which report enhanced responsivity values between 1.5 and 5 in all our devices with θ<1°, especially considering that our experiments are undertaken at finite temperatures (10K) and that the residual doping in our samples is typically ~$10^{11}$ cm$^{-2}$.

*Existence and size of energy gaps at the satellite Dirac points.* The presence of moiré potentials breaks the inversion symmetry of the heterostructure, which, in principle, could lead to a gap opening at both the main and the two satellite Dirac points[19,34,35]. Indeed, multiple electrical and optical measurements in the literature have provided convincing evidences of energy gaps present in aligned graphene/hBN devices at the main Dirac point (Δ) and the valence band satellite Dirac point ($Δ_h$)[2,20,22,36]. The size of these gaps is ~10-40 meV, with larger values occurring closer to perfect alignment (θ=0°)[3]. However, despite predictions[32,34,35,37,38], no energy gaps have been observed up to now in the case of the conduction band sDP ($Δ_e$).

We demonstrate that gate-dependent and zero-bias THz photodetection at different frequencies is a relevant technique to examine the presence of tiny ~meV global and local energy gaps at both satellite Dirac points $Δ_h$ and $Δ_e$ in graphene moiré superlattice systems. This is due to several reasons, including the fact that THz radiation covers an energy range comparable to the size of the gaps present in these moiré superlattices. Moreover, zero-bias, gate-dependent THz photocurrents have different (intra- or interband) origins and thus exhibit well-differentiated features depending on the chemical potential $E_F$ and the frequency of the incoming radiation.

For instance, at $E_F$ close to the valence band sDP (**Figure 3a**), device A shows a clear transition from an intraband to an interband type of photodetection when increasing the radiation frequency from 0.075 to 4.7 THz (see **Figure 3b**). Specifically, whereas the measured photocurrent responsivity $R_I(n)$ is consistent with an intraband-type mechanism (exhibiting a vanishing photocurrent and a sign change right at the hole sDP) for excitation frequencies equal to or below 2.5 THz (≈10 meV); $R_I(n)$ changes at higher frequencies and exhibits a pronounced and stable minimum right at the position of the sDP when the excitation frequency is 4.1 THz (≈17 meV), instead. The latter lineshape is a clear indication that interband transitions become dominant at frequencies ≥ 4.1 THz and provides a clear estimate of the energy gap size, $\Delta_h$ ≈17 meV. Such a gap value agrees well with the one extracted via traditional methods such as temperature dependent transport measurements, both in literature[2] and in our samples (see **Supporting Information Note 7**). On the other hand, the recorded zero-bias, gate-dependent THz photocurrents for Device B (**Figure 3c**) displays an intraband-type behavior over the measurement range (0.075-4.7 THz), which is attributed to the fact that $\Delta_h$ in more aligned devices is larger[2] than the highest radiation energy available in our experimental setup (~20 meV).

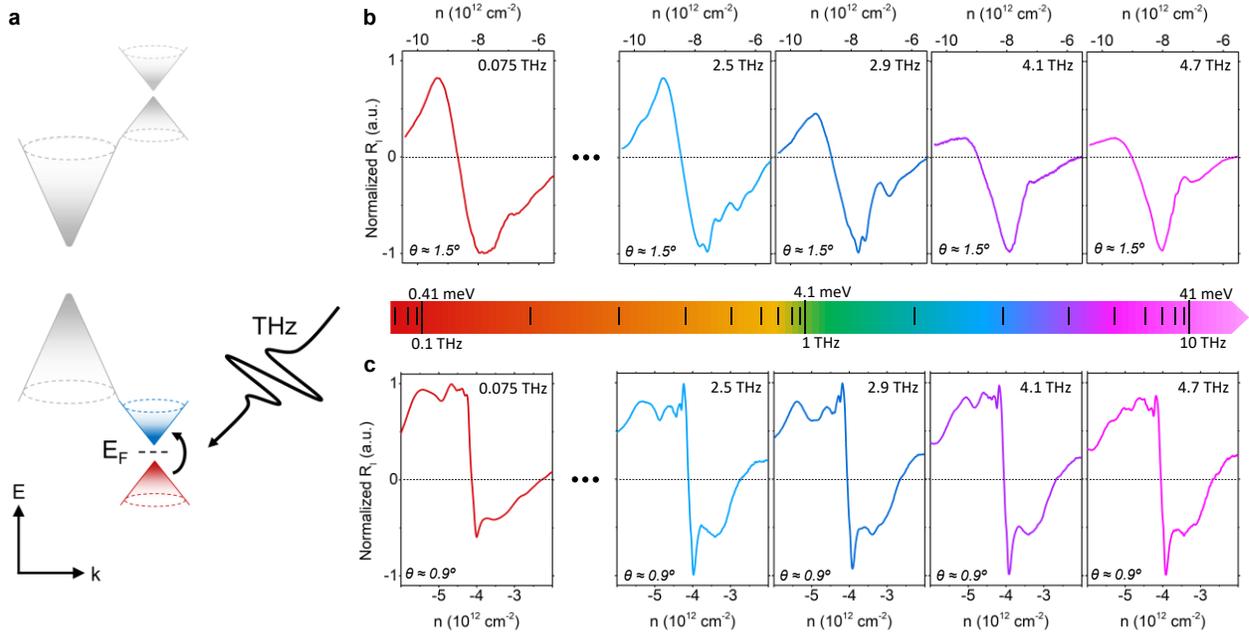

*Figure 3. Low temperature photocurrent spectroscopy measurements at the valence band sDP. **a**, Schematic bandstructure of a graphene/hBN heterostructure, highlighting the possible presence of interband transitions at the hole-band sDP at THz frequencies. **b**, Measured photocurrent*

*responsivity $R_I$ at different THz frequencies as a function of the carrier density, n, when the $E_F$ is close to the valence-band sDP for Device A. **c,** Measured photocurrent responsivity $R_I$ at different THz frequencies as a function of the carrier density, n, when the $E_F$ is close to the valence-band sDP for Device B. Measurements in panels **b** and **c** are undertaken at 10K.*

Importantly, gate-dependent THz photoresponses at multiple frequencies are also able to detect and assess the size of local energy gaps existing at the conduction band sDP ($\Delta_e$) in graphene moiré superlattices (Figure 4a). We remark that the existence of a non-zero gap $\Delta_e$ in graphene/hBN superlattices is an intriguing result which, despite predictions[32,34,35,37,38], has remained under debate up to now[2,19,20]. In fact, conventional (optical and electrical) probing techniques[2,20,22] have so far been unable to detect any gap at the conduction band sDP so far, and this is the main reason why several studies assume the absence of such energy gaps[2,20] (further analysis is provided in the **Discussion section**).   **Figure 4b** shows $R_I(n)$ measured close to the conduction-band sDP at different THz frequencies (from left to right $f$ = 0.075, 0.15, 0.3, 2.5 and 4.7 THz, respectively) in device A (θ~1.6˚). An interband-type photocurrent is present for frequencies $\geq$ 2.5 THz, therefore, the energy gap $\Delta_e$ in this device is estimated to be ≈10 meV.

In contrast, **Figure 4c** shows $R_I(n)$ measured close to the conduction-band sDP at different THz frequencies (from left to right $f$ = 0.075, 0.15, 0.3, 0.6 and 4.7 THz, respectively) in device B. Whereas a clear photoresponse of intraband type is measured at $f$ = 0.075 THz, a photoresponse of interband origin is already evident and stable at frequencies above 0.3 THz. This observation demonstrates the presence of a non-zero bandgap at the electron-band SDP in this device with θ~0.9° and suggests an approximate size of $\Delta_e \approx 1.2$ meV. By measuring and taking into account $R_I(n)$ close to the conduction band sDP in all devices under study (see **Supporting Information Note 8**), we corroborate that the size of $\Delta_e$ depends on the misalignment angle θ (see theoretical analysis below).

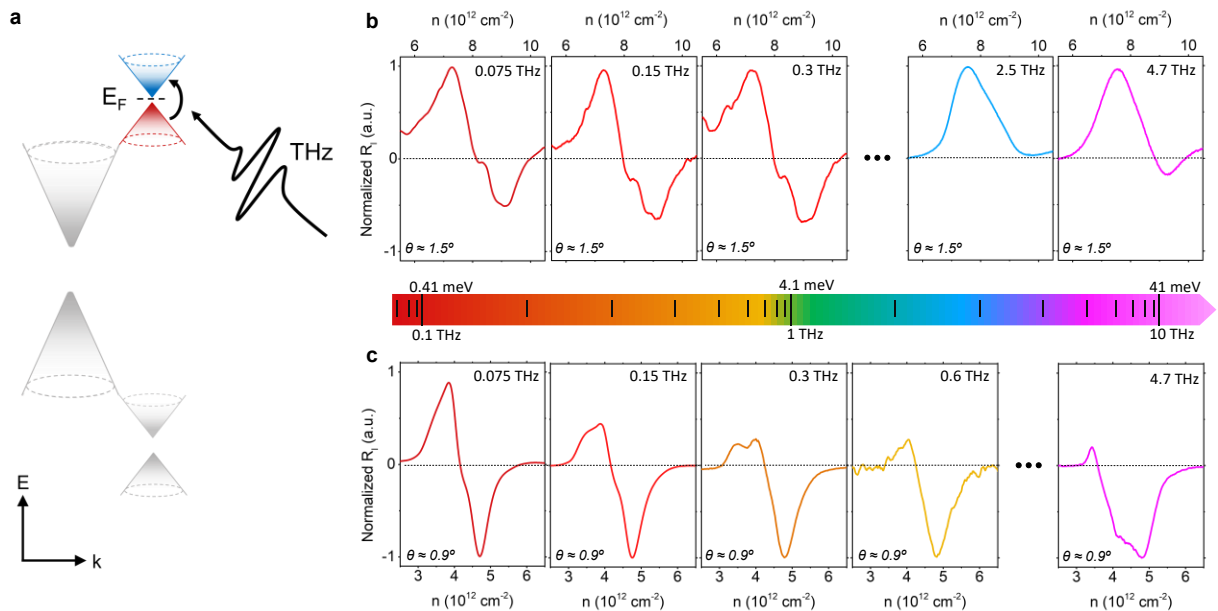

*Figure 4. Low temperature photocurrent spectroscopy measurements at the conduction band sDP. a, Schematic bandstructure of a graphene/hBN heterostructure, highlighting the possible presence of interband transitions at the electron-band sDP at THz frequencies. b, Measured photocurrent responsivity $R_I$ at different THz frequencies as a function of the carrier density, n, when the $E_F$ is close to the conduction-band sDP for device A. c, Measured photocurrent responsivity $R_I$ at different THz frequencies as a function of the carrier density, n, when $E_F$ is close to the conduction-band sDP for device B. Measurements in panels b and c are done at 10K.*

*Theoretical analysis.* The results and interpretation given above are fully supported by calculations of the band structure and first and second order conductivities of these systems with a tight-binding model (***Methods***). **Figure 5a** and **5b** depicts the simulated band structure of a representative graphene/hBN heterostructure near the valence and conduction band sDPs, respectively. The misalignment angle between graphene and hBN flakes in this simulation is set to θ=0.44°, and thus the sDPs induced by the superlattice potential occur around energies[1] $E_{SDP} = \pm\pi v_F^{(DP)}/\lambda_M \approx \pm 150$ meV (or equivalently, at carrier densities $n_{SP} \approx \pm 2.72\times10^{12}$ cm$^{-2}$), at the K, K' and/or Γ points of the Brillouin zone of the rhombohedral supercell used for the calculation (see details in ***Supporting Information Note 9***). Importantly, the band structure demonstrates that sDPs in both conduction and valence bands contain energy gaps, which are the result of Bragg scattering at the edges of the superlattice Brillouin zone[39]. In particular, the band splitting is rather large on the hole side, leading to an actual spectral gap $\Delta_h^1 > 20$ meV. We also highlight the fact that, similarly to other studies reported in literature[35,38], **Figure 5b** does not display a global band gap at the conduction band sDPs. Instead, we observe a complicated band texture, including a series of avoided band crossings, giving rise to several local energy gaps (marked $\Delta_e^1$ to $\Delta_e^4$) between consecutive bands. The smallest of these is ~ 3 meV, an order of magnitude smaller than the hole side gap and in good agreement with measurements shown in Figure 4 for samples with θ < 1degree. For clarity, **Figure 5c** displays the evolution of all calculated energy gaps in the system ($\Delta_h^1, \Delta_e^1, \Delta_e^2, \Delta_e^3, \Delta_e^4$) for a range of twist angles θ between 0 and 1.8° as well as the energy gaps extracted from our devices via THz photocurrent measurements ($\Delta_h^{exp}, \Delta_e^{exp}$). In general, the calculated energy gaps are in good agreement with the values extracted from all our devices and measurements (**Figures 3, 4** and ***Supporting Information Note 8***), particularly given that any residual strain and/or twist-angle variations present in experimental samples will alter (reduce) the magnitude of gaps at the sDPs[37].

Moreover, we highlight that the evolution of the size of the local energy gaps in the conduction-band sDPs when increasing θ is non-monotonic. This behavior is different from the simpler evolution of the size of energy gaps in the valence-band sDP with θ ($\Delta_h^1$ decreases for larger θ).

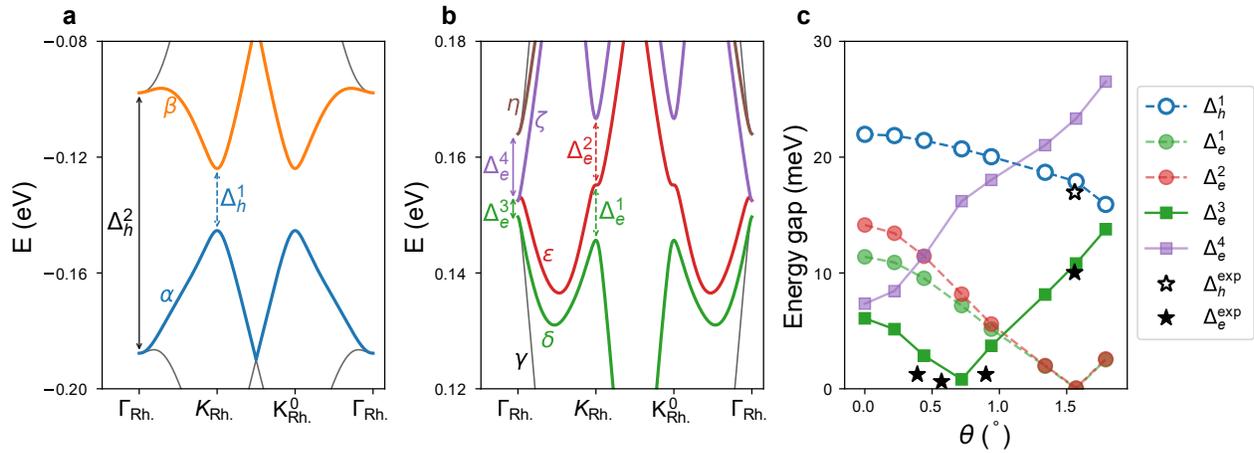

*Figure 5. Band structure calculations of graphene/hBN moiré superlattice systems.* ***a,*** *Calculated bandstructure of a graphene/hBN heterostructure, highlighting the existence of energy gaps $\Delta_h^1$ at the hole-band sDPs with sizes around tens of meV.* ***b,*** *Calculated bandstructure of a graphene/hBN heterostructure, highlighting the existence of energy gaps at the electron-band sDPs (marked $\Delta_e^1$ to $\Delta_e^4$) with sizes around a few meV.* ***c,*** *Dependence of the energy gap sizes at the valence and conduction band sDPs on the misalignment angle θ. Experimental points (star symbols) represent the minimum energy gap values at the electron-band sDP ($\Delta_e^{exp}$) or the hole-band sDP ($\Delta_h^{exp}$), extracted via THz photocurrent spectroscopy.*

The existence of energy gaps at the sDPs with energy scales on the order of ~meV promotes optical transitions at THz frequencies due to enhanced joint density of states (JDOS) related to the neighboring moiré minibands (see JDOS calculations in *Supporting Information Note 9*). These optical transitions are vertical (i.e. direct) and occur at specific **k**-points in the reciprocal space, given the fact that photoexcitation processes satisfy momentum conservation and THz photons have a very small momentum with respect to the dimension of the Brillouin zone of the system. In consequence, as shown in **Figures 6a** and **6b**, the interband optical conductivity of graphene moiré superlattices shows a notable spectral weight at doping levels $E_F = \pm E_{sDP}$, when the incoming energy of the THz radiation $\hbar\omega$ exceeds the distinct energy gap(s) present in the system (see annotations $\Delta_h^1, \Delta_e^1, \Delta_e^2, \Delta_e^3, \Delta_e^4$ in the two panels, which correspond to the energy levels highlighted in **Figures 5 a,b**). In more detail, the interband activity at the hole band sDP (i.e. doping level $E_F = -E_{sDP}$, **Figure 6a**) acquires non-zero values at energies larger than $\Delta_h^1$, and thus covers part of the THz energy range. In contrast, due to the presence of smaller gaps, the interband activity in samples with θ < 1° is appreciable in most of the THz range (i.e. energies > $\Delta_e^3$ ~ 3meV) at the conduction band sDP (doping level $E_F = +E_{sDP}$, **Figure 6b**). All these trends are in close agreement with the overall photoresponse measured around the valence and conduction band sDPs at different THz frequencies in our devices (**Figures 3, 4** and *Supporting Information Note 8*).

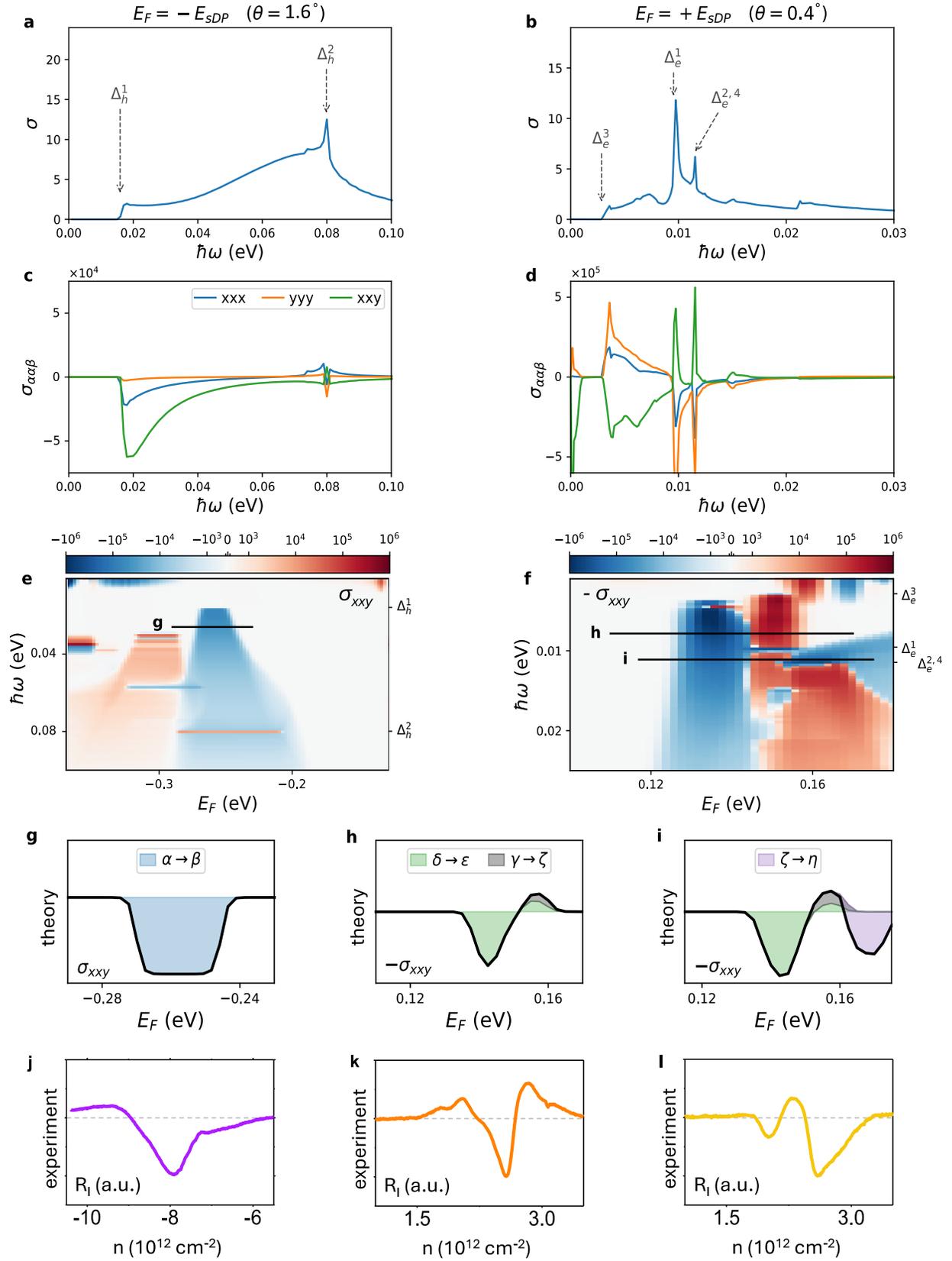

*Figure 6. First and second order conductivity calculations of graphene/hBN moiré superlattice systems. Comparison with experiments. a,* Interband conductivity of graphene/hBN moiré superlattices as a function of the excitation energy $\hbar\omega$ at the position of the sDP in the valence band (doping level $E_F = -E_{sDP}$) for a simulated sample with $\theta=1.6°$. *b,* Interband conductivity of graphene/hBN moiré superlattices as a function of the excitation energy $\hbar\omega$ at the position of the sDPs in the conduction band (doping level $E_F = +E_{sDP}$) for a simulated sample with $\theta = 0.4°$. *c,* Dependence of the shift conductivity components as a function of the excitation energy $\hbar\omega$ at the position of the sDP in the valence band (doping level $E_F = -E_{sDP}$) for the simulated sample with $\theta=1.6°$. *d,* Same than panel *c* for the conduction band (doping level $E_F = +E_{sDP}$) for the simulated sample with $\theta = 0.4°$. *e,f* Maps of the shift conductivity component $\sigma_{xxy}$ as a function of the excitation energy $\hbar\omega$ and the doping level $E_F$ around the position of the sDPs. In the valence band for the simulated sample with $\theta=1.6°$, panel *e*, and in the conduction band for the simulated sample with $\theta = 0.4°$, panel *f*. *g,* Linecut of the calculated $\sigma_{xxy}(E_F)$ in panel *e* around the valence band sDP for an excitation energy $\hbar\omega$ larger than $\Delta_h^1$. *h* and *i,* Linecuts of the calculated $\sigma_{xxy}(E_F)$ in panel *f* around conduction band sDPs for excitation energies $\hbar\omega$ larger than $\Delta_e^3$ and $\Delta_e^1$, respectively. *j,* Measured $R_I$ of device A ($\theta\approx 1.6°$) around the valence band sDP for an excitation frequency 4.1 THz. *k* and *l,* Measured $R_I$ of device D ($\theta\approx 0.4°$) around the conduction band sDPs for excitation frequencies 0.3 and 0.6 THz, respectively. Dotted gray lines in panels *j-l* correspond to $R_I=0$.

<u>*Quantum geometric photocurrents.*</u> For completeness, we now consider the microscopic mechanism giving rise to a robust zero-source-drain bias photocurrent in our graphene moiré superlattice systems for excitation energies larger than the energy gaps present at the valence and conduction band sDPs. We remark that broken inversion symmetry and a local gap opening at an avoided crossing of Bloch bands induces a finite Berry curvature in the bands immediately above and below the energy gap. This geometric property is a consequence of Bragg scattering from the moiré superlattice, which mixes the pseudospin textures in the system[39]. Furthermore, the peaked $R_I(n)$ observed at the doping levels $\pm n_{SP}$ cannot be explained by intraband activity but it is more characteristic of intraband transitions. Given also the fact that the measured photoresponse arises from linearly polarized light, is highly sensitive to the atomic configuration of the graphene/hBN superlattice (i.e. to the misalignment angle θ) as well as the frequency of the incoming THz radiation, one can reasonably assign its origin to quantum geometric shift photocurrents occurring in van der Waals heterostructures with broken inversion symmetry[17]. Such phenomena are second-order responses which arise as a result of a real-space displacement experienced by an electron wave-function upon an optical transition and have been recently observed in moiré systems such as twisted bilayer graphene around the main DP[10,40].

The dependence of three components of the shift conductivity ($\sigma_{xxx}$, $\sigma_{yyy}$, $\sigma_{xxy}$) with the excitation energy $\hbar\omega$ at the two selected chemical potentials in the valence and conduction minibands, $E_F = -E_{SDP}$ and $E_F = +E_{SDP}$, is shown in **Figures 6c** and **6d** respectively. The induced photocurrent in a real system will contain several contributions, each proportional to a different shift conductivity component, with their relative weights dictated by the exact alignment of the device geometry and the polarization of the THz radiation[17]. In general, all components of the shift conductivities become non-zero at excitation energies exceeding the energy gaps of the valence and conduction band sDPs ($\hbar\omega > \Delta_h, \Delta_e$), with all three $\sigma_{xxx}, \sigma_{yyy}, \sigma_{xxy}$ showing prominent peaks at these energies. $\sigma_{xxy}$, which has the largest magnitude of the components shown, will now be used as a representative example to further analyze the shift photocurrent behavior near the valence and conduction band sDPs. Note that we may also invert its sign, which has the effect of reversing the arbitrary positive direction of one of the spatial directions, for an easier comparison with experimental results.

**Figure 6e** and **6f** maps the evolution of σ$_{xxy}$ as a function of $E_F$ and $\hbar\omega$ close to the valence and conduction sDPs, respectively. Around the valence band sDP (**Figure 6e**), the largest shift current occurs when the chemical potential is placed within the energy gap (position $E_F = -E_{sDP}$). Getting away from the sDP, σ$_{xxy}$ diminishes and eventually vanishes due to Pauli blocking[17]. An equivalent situation occurs in the vicinity of the conduction band sDP (**Figure 6f**). The main difference lies in the fact that, due to the more convoluted band structure existing at these energies (**Figure 5b**), the dependence of the calculated shift photocurrent on $E_F$ may exhibit several sign changes and relative maxima at different excitation energies $\hbar\omega$. This can be seen more clearly by examining horizontal cuts taken along the black lines in **Figures 6e** and **6f** which are shown in panels **g-i**. The valence sDP minimum (**Figure 6g**) is relatively simple and can be exactly identified with transitions between bands $\alpha$ and $\beta$ in (**Figures 5a and S11**), once the excitation energy $\hbar\omega > \Delta_h^1$ (the minimum energy gap separating these bands). The conduction band cuts display a more complex behavior around $E_F = +E_{sDP}$, as even contributions from the same two bands can have opposite signs. This is shown in **Figure 6h** for transitions between the bands labelled $\delta$ and $\varepsilon$ in **Figure 5b**, where transitions at lower $E_F$ (i.e., those at $K_{Rh.}$, requiring $\hbar\omega > \Delta_e^1$) have an opposite sign to those at slightly higher $E_F$ (i.e., at $\Gamma_{Rh.}$, requiring $\hbar\omega > \Delta_e^3$). Increasing the excitation energy further allows transitions between other bands, as demonstrated in **Figure 6i**, where the additional contribution of transitions between bands $\zeta$ and $\eta$ give rise to an additional sign change and minima. Intriguingly, as shown in **Figures 6j**, **6k** and **6l**, we observe similar complex interband photocurrent patterns in the measured photoresponse of some of our experimental devices. Specifically, the measured response in device A ($\theta \approx 1.56°$) at 4.1 THz (**Figure 6j**) shows a photocurrent $R_I(n)$ with a single wide minimum occurring at a carrier density $n \approx -8 \times 10^{12}$ cm$^{-2}$ (i.e. fermi level $E_F \approx -0.27$ eV), which is indicative of a transition between two bands separated by a rather large gap existing at the valence band sDP. Both the line shape and the Fermi level position of the minimum agree well with the calculated $\sigma_{xxy}(E_F)$ for a simulated device with $\theta = 1.6°$ (**Figure 6g**). Meanwhile, multi-peak structures occuring at Fermi levels between 0.14-0.16 eV are observed (**Figures 6k** and **6l**) for device D ($\theta \approx 0.4°$) at 0.3 and 0.6 THz, respectively. Such features closely match the curves $\sigma_{xxy}(E_F)$ simulated for a corresponding device with $\theta = 0.4°$ (**Figure 6h** and **6i**), and are thus indicative of the rich interband transition landscape provided by the various subbands present at the conduction band sDP.

*Comparison with other techniques.* The presented results prompt gate-tunable THz photocurrent spectroscopy as a robust and comprehensive technique to probe several features of the miniband electronic structure of graphene in the presence of moiré potentials, including Fermi surface parameters, marked energy levels and subtle quantum geometry fingerprints of these electronic states. This is because the measured photoresponses show well defined dependences (on both excitation energy and the chemical potential) when electrons transition within the same energy band or between the different bands present in these systems, and hence the technique is sensitive to the distinct electronic features and band parameters.

An intraband-type of photocurrent occurs at the lowest THz frequencies and depends on details of the electronic Fermi surface of the system. In particular, as shown in the former sections, intraband photoresponses are sensitive to the overall density of states $DOS$ of the system and thus, to relevant miniband parameters such the valley degeneracy $g_v^{(DP)}$ and the Fermi velocity $v_F^{(sDP)}$ of charge carriers at the satellite Dirac points. Interestingly, one can extract these parameters in a simple way from the recorded THz photoresponse $R_I(n)$, only by examining the ratio of the responsivity $R_I$ measured around the main and satellite Dirac points $\delta R_I^{sDP}/\delta R_I^{DP}$ (Eq.1). More conventional techniques reported in literature which are sensitive to $g_v^{(DP)}$ and $v_F^{(sDP)}$ such as scanning tunnelling microscopy (STM)[1] or capacitance spectroscopy[33] measurements are able to disclose this information in a more convoluted way, by monitoring (and fitting) the evolution the $DOS$ of the system with respect to the applied gate-voltage or device temperature, respectively.

Moreover, the lack of inversion symmetry existing in these moiré materials results in the presence of tiny (< 20 meV) energy gaps at both valence and conduction bands sDPs as well as the generation of quantum geometric shift (bulk) photocurrents when the excitation energy of the incoming radiation is larger than the size of these energy gaps. The presence of this distinct type of interband photocurrent allows to probe key miniband parameters including the actual size of the existing bandgap at the valence band sDP ($\Delta_h$) as well as the more complicated quantum texture present at the conduction band sDPs, featuring a number of avoided band crossings which give rise to a series of local energy gaps ($\Delta_e$) between consecutive bands (**Figure 5b**). Intriguingly, some of the extracted parameters (e.g. energy gaps $\Delta_e$) are undetectable by alternative techniques reported in literature. The absence of a well-defined gap (i.e. a band gap) in the conduction band sDP (or equivalently, the existence of a Fermi surface and an intraband contribution[35] at these energies) impedes the observation of a non-zero $\Delta_e$ via commonly used techniques based on electrical measurements, including temperature dependent transport[3], tunnelling spectroscopy[20] or capacitance spectroscopy[33] (for example, see transport data and subsequent analysis of some of our devices in the Supporting Note 7). In addition, well-known optical techniques utilized to directly probe bandstructure of graphene-based superlattices such as micro- or nanometer scale angle-resolved photoemission spectroscopy (ARPES) have a limited energy resolution of ~30 meV[22] and thus are also not capable of resolving the tiny energy gaps < 20 meV present at the conduction band sDPs.

We remark that gate-dependent THz photoresponse of an interband origin are not subjected to any of the two former limitations. On the other hand, photoexcitation processes occurring at the sDPs in electron- and hole-bands satisfy both momentum and energy conservation. For momentum conservation, the generation of electron-hole pairs is restricted to occur between states with the same wave vector value *k* in reciprocal space, owing to the very small momentum of the incident THz photons. In this sense, the measured bulk (interband) photocurrents are a direct consequence of vertical optical transitions which exclusively take place at specific **k**-points in the vicinity of the sDPs and hence are extraordinarily sensitive to the size of energy gaps occurring in the minibands of graphene-based moiré superlattice systems. Regions localized away from the sDPs, including un-gapped regions as a result of e.g. shunted edges[40,41], behave only as current collecting leads.

Finally, from a practical point of view, our measurements demonstrate that a conservative estimate of the energy resolution offered by gate-dependent THz photoresponse measurements is well-below 1meV. This is clearly shown in **Figure 5c** (main text) and in **Figure S10c** of Supporting Note 8, which display interband transitions at radiation energies ~0.6 meV (0.15THz). Such level of resolution is possible thanks to the notably distinct gate- and frequency-dependence exhibited by the different physical mechanisms giving rise to zero drain-bias THz photocurrents:

whereas the optoelectronic response of interband origin (bulk photocurrents) is maximal at the sDPs, intraband photoresponses vanish at these energies. A detailed benchmarking between gate-dependent, THz photocurrent spectroscopy measurements and more conventional techniques to probe the different electronic parameters of graphene samples subjected to moiré superlattice potentials is depicted in **Table 1**.

*Table 1. Comparison between different techniques able to probe electronic parameters of the miniband structure of graphene/hBN moiré superlattices and reported energy resolution.*

|  | *Miniband structure parameters and energy resolution* | | | | |
|---|---|---|---|---|---|
| ***Probing technique*** | $\Delta_h$ [meV] | $\Delta_e$ [meV] | Probe $g_v^{(sDP)}$? | Probe $v_F^{(sDP)}$? | Energy resolution [meV] |
| *Transport measurements (Ref.[3])* | tens of meV (angle dependent) | NO [*] | NO | NO | ~ 1-2 [***] |
| *Tunneling spectroscopy (Ref.[20])* | tens of meV (angle dependent) | NO [*] | NO | NO | ~ 3 |
| *Capacitance spectroscopy (Ref.[33])* | NO [**] | NO [*] | YES | YES | ~ 5 |
| *STM (Ref.[2])* | NO [**] | NO [*] | YES | YES | - |
| *ARPES (Ref.[27])* | 100 meV | NO [**] | NO | NO [**] | ~ 30 |
| *THz photocurrent spectroscopy (this work)* | tens of meV (angle dependent) | 0.6-12 meV (angle dependent) | YES | YES | <0.6 [****] |

*(\*)* *Parameter not reported in literature since the technique is sensitive to the overall bandgap of the system.*
*(\*\*)* *Parameter not reported in literature due to the limited energy resolution offered by the technique.*
*(\*\*\*)* *Uncertainty in the extracted gaps are set by the determination of the linear (thermally activated) regime for the fit[2].*
*(\*\*\*\*)* *Conservative value estimated from the minimum gap size demonstrated in the present work. This value is given by the minimum energy at which we observe an interband type of photocurrent in our samples (0.15THz, see Figure S10c in Supporting Note 8). Measurement precision could eventually improve if additional low-energy lines would be available in the set-up.*

# CONCLUSIONS

We have shown gate-tunable THz photocurrent spectroscopy as a robust and comprehensive technique to probe several features of the miniband electronic structure of graphene in the presence of moiré potentials, including Fermi surface parameters, as well as intriguing energy levels and band textures not previously detected. On top of that, by observing bulk photocurrents in graphene moiré superlattices, our study demonstrates THz photocurrent measurements are also able to capture quantum geometric properties of electron states in the minibands of these materials (changes in the internal structure of electron wavefunctions between quantum states coupled by electromagnetic fields).

Finally, from a more technological perspective, we stress that the enhanced zero drain-bias responsivity values $\delta R_I^{sDP}/\delta R_I^{DP}$ appearing close to the sDPs in graphene/hBN moiré detectors with θ <1° are subsequently accompanied by a lower noise equivalent power ($NEP$) at the satellite Dirac points $NEP^{sDP}$ with respect to the main Dirac point $NEP^{DP}$ (see **Supporting Information Note 10**). In particular, $NEP$ ratios $NEP^{sDP}/NEP^{DP}$ reach values down to 0.2 in these devices. As such, these two relevant device parameters (enhanced responsivity $\delta R_I^{sDP}/\delta R_I^{DP}$ and reduced

*NEP* close to the satellite Dirac points) further demonstrate graphene/hBN moirè superlattices with alignment angles θ < 1˚ as convenient materials for sensitive and low noise THz detection.

# METHODS

**Device fabrication:** Our aligned devices with angles θ between 0.4˚ and 1.6˚ are made of monolayer or bilayer graphene encapsulated between hBN flakes. Graphene and hBN flakes were first mechanically exfoliated and identified using Raman spectroscopy. Bottom and top hBN flakes were chosen with similar thicknesses (between 20 nm and 30 nm) and identified via optical contrast and profilometer measurements. We chose elongated graphene and hBN flakes with straight edges to identify the crystallographic orientation of the graphene and hBN crystals. The stacking process of the moiré heterostructures was made using a dry transfer technique with a polypropylene carbonate film on a polydimethylsiloxane stamp. During the assembling process, we intentionally aligned close to 0º the straight edges of the top hBN and the MLG or BLG flakes (see further details in **Supporting Information Note 1**). The resulting stacked MLG/hBN moiré heterostructures were additionally characterized via Raman spectroscopy (see **Supporting Information Note 3**).

We then contacted all our samples with one-dimensional electrical contacts. For this, the heterostructures were patterned using electron beam lithography (EBL) to define contacts areas using PMMA (6% in chlorobenzene) as resist. Subsequently, the heterostructures were dry-etched in an ICP-RIE Plasma Pro Cobra 100 in a $SF_6$ atmosphere (40 sccm, pressure 6 mTorr, power 75 W at 10 °C) followed by an e-beam evaporation process at very low pressure (< $5\times10^{-8}$ Torr) of 3.5 nm of Cr and 65 nm of Au to deposit the metallic layers.

Afterwards, two distinct processes were undertaken to shape and finalize our samples, depending on the device architecture. Top gates in SC and IDGT devices (samples A, B, D and E) are fabricated via EBL and e-beam evaporation of 5 nm Cr and 45 nm Au. Meanwhile, the definition of the MC geometry was undertaken via additional EBL and dry-etching processes.

**Electrical measurements:** Transport measurements were carried out via standard lock-in technique where a pseudo-dc current (11.33 Hz) of 10 nA was injected into the drain and then collected in the source. The generated voltage drop was recorded by a lock-in amplifier (SR860). Measurements on devices A, B, D and E were undertaken via a two-terminal configuration while device C was characterized via a four-terminal configuration. The carrier density in the channel is controlled by the voltage applied either to the back-gates $V_{BG}$ or top gates $V_{TG}$ using a voltage generator (Keithley 2614B). For the back-gate bias, we directly applied a bias to the highly doped Si substrate with a typical range from -60V to 60 V (for $SiO_2$ thickness of ~300 nm). For the case of top-gate bias, values typically ranged from -4V to 4V (for hBN thicknesses of ~20 nm).

**THz photoresponse measurements:** All our graphene-based moiré superlattice devices were placed inside a cryostat system (ARS µDrift Nanoscience Cryostat) with optical access to perform the measurements. In our experiment, we collected the THz photocurrent through the source and drain electrodes at zero bias. For low-frequencies, the measurements were performed using a THz source based on Schottky diodes multipliers chains (TeraSchottky from Lytid) to generate output

THz linear polarized frequencies at 0.075 THz (optical output power of ~50 mW), 0.15 THz (optical output power of ~30 mW), 0.3 THz (optical output power of ~6 mW) and 0.6 THz (optical output power of ~0.3 mW). For high-frequencies, measurements were performed with a quantum cascade continuous-wave laser (TeraCascade 2000 series from Lytid) with output linearly polarized lines at 2.5 THz (optical output power of ~3.2 mW), 2.9 THz (optical output power of ~2.1 mW), 4.1 THz (optical output power of ~3.2 mW) and 4.7 THz (optical output power of ~4.8 mW). The output THz radiation, electrically modulated at 667 Hz, was collimated with an off-axis gold parabolic mirror and finally focused on the moiré THz devices by using a THz lens made of TPX (Polymethylpentene) with a focal distance of 100 mm. Linearly polarized THz radiation was used to undertake all measurements. The photocurrent generated in the device channel was collected in the drain contact and then amplified and measured using a low noise current preamplifier (Stanford Research, SR570) in series with a lock-in amplifier (SR860). Carrier density on the channel was controlled by applying a voltage on the back-gate or top-gate electrodes using a voltage generator (Keithley 2614B). Source terminals were kept grounded during the photocurrent experiments.

**Tight binding and shift conductivity calculations:** Periodic twisted graphene/hBN structures were first generated by finding superlattice vectors in each of the untwisted layers which had almost identical length and a relative rotation similar to the desired twisted angle. Periodic twisted graphene/hBN structures were first generated by finding two lattice vectors, one from each of the untwisted layers, which have almost identical length and a relative rotation similar to the desired twist angle. These two vectors were then perfectly aligned by rotating the hBN layer and applying a miniscule strain (of magnitude 0.01% or less), and the resultant vector gives one of the superlattice vectors of a rhombohedral unit cell, with the second superlattice vector given by a $\frac{\pi}{3}$ rotation.

The 0.44°-rotated system shown in the main text contains 10424 atoms in its unit cell and its hBN layer is subject to a biaxial strain $\varepsilon = 1.9 \times 10^{-5}$. The electronic properties of the resulting twisted, multi-layered graphene / hBN systems are described using a tight-binding Hamiltonian $\widehat{H} = \sum_i \epsilon_i c_i^\dagger c_i - t(\vec{d}) \sum_{\langle i,j \rangle} (c_i^\dagger c_j + c_j^\dagger c_i)$, where $c_i^\dagger$ and $c_i$ are the creation and annihilation operators for an electron in the $p_z$ orbital at site i. The onsite energies $\epsilon_i$ depend on the atomic species, with $\epsilon_C = 0.0$ eV, $\epsilon_B = 3.34$ eV and $\epsilon_N = -1.4$ eV. The hopping parameter $t(\vec{d})$ depends on the distance vector between both sites and is given by $-t(\vec{d}) = V_{pp\pi}(d) \left[1 - \left(\frac{\vec{d} \cdot \vec{e_z}}{d}\right)^2\right] + V_{pp\sigma}(d) \left(\frac{\vec{d} \cdot \vec{e_z}}{d}\right)^2$, and its two components give both the in-plane and out-of-plane contributions to the hopping. The transfer integrals $V_{pp\pi}(d)$ and $V_{pp\sigma}(d)$ can be written in terms of their (unstrained) graphene values as $V_{pp\pi}(d) = V_{pp\pi}^0 e^{-\frac{d-a_0}{r_0}}$ and $V_{pp\sigma}(d) = V_{pp\sigma}^0 e^{-\frac{d-d_0}{r_0}}$, with $V_{pp\pi}^0 = -2.7\ eV$ and $V_{pp\sigma}^0 = 0.48\ eV$. $a_0 = 1.42$ Å and $d_0 = 3.35$ Å are the corresponding unstrained in-plane and out-of-plane atomic separations, $r_0 = 0.453$ Å is a decay length, and $\vec{e_z}$ is the unit vector in the out-of-plane direction. We include all hopping terms whose corresponding separation in the xy-plane is less than a cutoff of 2.94 Å.

The band structures in **Figure 5** are obtained by directly solving for the eigenvalues of the Bloch Hamiltonian. The optical and shift conductivities require the calculation of matrix elements[18,42,43] of the form $h_{ab}^\alpha = \langle a | \nabla_{k_\alpha} H | b \rangle$, where $| a \rangle, | b \rangle$ are eigenstates of the Hamiltonian and $\alpha = x, y, z$. To calculate the optical conductivity, we follow the approach in Ref.42, which averages the x and y components. We also employ the improved triangular method introduced in Ref.42 to replace the integration over the irreducible Brillouin zone (IBZ) with a sum of contributions from individual triangular sections into which it is divided, as this approach typically requires a sparser sampling of k-space. The results for the 0.44°-twisted system use 2145 k-points in the IBZ.

We follow Ref.18 in our simulation of the shift current response, which is described by a rank-three tensor $\sigma_{\alpha\alpha\mu}$, where $\alpha, \mu$ = x,y respectively indicate the spatial components of the induced currents and electric fields. We do not consider $\alpha, \mu$ = z in this work due to the in-plane polarization of the radiation but note it has been considered in other works[43]. The $\sigma_{\alpha\alpha\mu}$ results shown in the main text are calculated following Eqs. (10) and (11) in Ref 18, where the k-space integrations are once more performed using the improved triangular method from Ref.42. The different pairwise contributions, shown in **Figures 6 g, h** and **i** are calculated by considering individual terms from the summation in Eq (10) of Ref.18

# AUTHOR INFORMATION

**Author Contributions**

J.A.D-N. and J.M.C. conceived the idea, designed and built the experiments, fabricated the samples, performed the measurements and analyzed the data. S.R.P. carried out the numerical simulations. D.V. provided technical assistance during the experimental characterization. Y.M.M. and J.E.V-P. conceived the experimental setup used in the study. K.W. and T.T. fabricated the hBN crystals. W.K, M.P., J.L.C. and P.A-G. helped in the interpretation of the results. The the manuscript was written by J.A.D-N. and J.M.C., with input from all authors. All authors have given approval to the final version of the manuscript.

# ACKNOWLEDGEMENTS

The authors thank the support from the Ministry of Science and Innovation (MCIN) and the Spanish State Research Agency (AEI) under the grants PID2021-128154NA-I00, PID2021-126483OB-I00, PID2022-141304NB-I00 and CNS2024-154588 funded by MCIN/AEI and by "ERDF A way of making Europe". This work has been also supported by Junta de Castilla y León co-funded by FEDER under the Research Grant number SA103P23. Financial support of the Department of Education of the Junta de Castilla y León and FEDER Funds is also gratefully acknowledged (Escalera de Excelencia CLU-2023-1-02). J.A.D.N. thanks for the support from the Universidad de Salamanca for the María Zambrano postdoctoral grant funded by the Next Generation EU Funding for the Requalification of the Spanish University System 2021–23, Spanish Ministry of Universities. J.A.D.N and J.M.C. acknowledge financial support by the MCIN and AEI "Ramón y Cajal" program (RYC2023-044965-I and RYC2019-028443-I) funded by MCIN/AEI and by "ESF Investing in Your Future". J.M.C., P.A.G. and W.K. also acknowledge financial support from the European Union (ERC StG CHIROTRONICS ID 101039754, ERC CoG TWISTOPTICS ID 101044461 and ERC AdG TERAPLASM ID 101053716, respectively). Views and opinions expressed are however those of the author(s) only and do not necessarily reflect those of the European Union or the European Research Council. Neither the European Union nor the granting authority can be held responsible for them. WK acknowledges also the support of "Center for Terahertz Research and Applications (CENTERA2)" project (FENG.02.01-IP.05-T004/23) carried out within the "International Research Agendas" program of the Foundation for Polish Science co-financed by the European Union under European Funds for a Smart Economy Program. S.R.P. wishes to acknowledge funding from the Irish Research Council under the Laureate awards programme, and the gazelle computational facility in the School of Physical Sciences at DCU, which is supported by Intel Ireland. M.P. acknowledges funding from the European Union NextGenerationEU/PRTR project Consolidación Investigadora CNS2022-136025. K.W. and T.T. acknowledge support from the JSPS KAKENHI (Grant Numbers 21H05233 and 23H02052) , the CREST (JPMJCR24A5), JST and World Premier International Research Center Initiative (WPI), MEXT, Japan. Authors also acknowledge USAL-NANOLAB for the use of Terahertz and Clean Room facilities.

# Supporting Information for

# "Unveiling the Miniband Structure of Graphene Moiré Superlattices via Gate-dependent Terahertz Photocurrent Spectroscopy"

Juan A. Delgado-Notario[1*], Stephen R. Power[2], Wojciech Knap[3,4], Manuel Pino[5], Jin Luo Cheng[6], Daniel Vaquero[7], Takashi Taniguchi[8], Kenji Watanabe[8], Jesús E. Velázquez-Pérez[1], Yahya Moubarak Meziani[1], Pablo Alonso-González[9], José M. Caridad[1,10*]

[1]Departamento de Física Aplicada, Universidad de Salamanca, 37008 Salamanca, Spain

[2]School of Physical Sciences, Dublin City University, Glasnevin, Dublin 9, Ireland

[3]CENTERA Labs, Institute of High Pressure Physics, Polish Academy of Sciences, Warsaw 01-142, Poland.

[4]Centre for Advanced Materials and Technologies CEZAMAT, Warsaw University of Technology, Warsaw 02-822, Poland.

[5]Departamento de Física Fundamental (IUFFyM) y GIR Nanotecnología, Universidad de Salamanca, 37008 Salamanca, Spain

[6]GPL Photonics Laboratory, State Key Laboratory of Luminescence and Applications, Changchun Institute of Optics, Fine Mechanics and Physics, Chinese Academy of Sciences,

Changchun, Jilin 130033, People's Republic of China and University of Chinese Academy of Sciences, Beijing 100039, China

[7]Zernike Institute for Advanced Materials, University of Groningen, 9747 AG Groningen, The Netherlands

[8]Research Center for Electronic and Optical Materials, National Institute for Materials Science, Tsukuba305-0044, Japan

[9]Department of Physics, University of Oviedo, Oviedo 33006, Spain

[10]Unidad de Excelencia en Luz y Materia Estructurada (LUMES), Universidad de Salamanca, Salamanca 37008, Spain

*  *emails*:   juanandn@usal.es,  jose.caridad@usal.es

# Note 1 – Additional fabrication details.

All our devices are fabricated on highly doped Si substrates (acting as back-gate) with 300 nm thermally growth $SiO_2$ on top. To assemble the aligned moiré heterostructures (rotation angle θ < 2˚) we intentionally select elongated flakes with constant width and straight edges (**Figures S1 a and b**). This helps us to identify the crystallographic orientation of hBN and graphene crystals and therefore ensure a precise alignment between flakes in the heterostructure (**Figure S1 c**). We further highlight that the use of high temperatures (up to 180 ºC) during the encapsulation process facilitates both the fabrication of clean interfaces[1] and most importantly, the final crystallographic alignment close to 0˚ of graphene and hBN crystals.

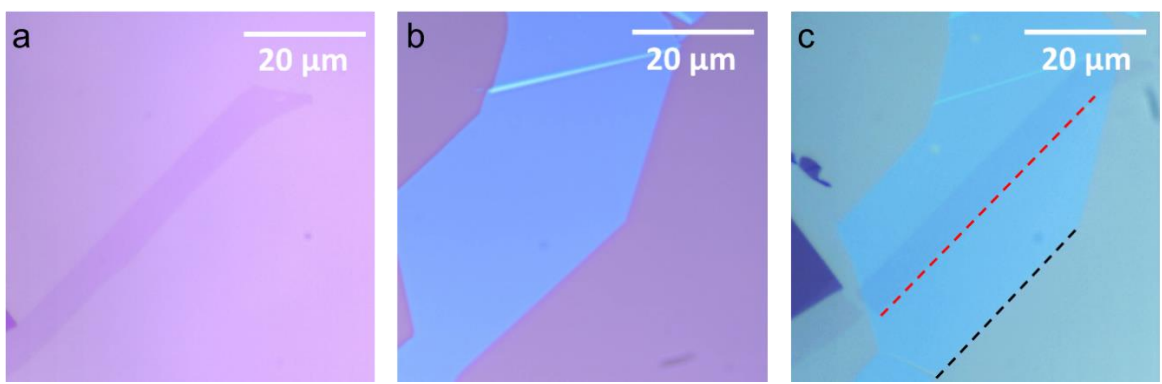

**Figure S1. Stacking process of graphene moiré heterostructures.** Optical image of the exfoliated flakes of **a,** Graphene, **b,** hBN and **c,** the fabricated half-heterostructure with the aligned top hBN and graphene flakes. Dashed lines indicate the crystallographic orientation of these crystals to guide the eye.

**Table S1** shows specific details of all moiré superlattice devices fabricated in this work, including the use of monolayer (MLG) or bilayer (BLG) graphene

| Device | Crystals | Top hBN Thickness (nm) | Bottom hBN Thickness (nm) | Device length $L_{CH}$ (μm) |
|---|---|---|---|---|
| A | MLG and hBN | 28 | 32 | 6 |
| B | MLG and hBN | 16 | 22 | 3 |
| C | MLG and hBN | 22 | 25 | 8.5 |
| D | MLG and hBN | 21 | 25 | 23.5 |
| E | BLG and hBN | 20 | 24 | 23.5 |

**Table S1.** Details of all graphene based heterostructures used in this work.

# Note 2 – Additional graphene moiré THz devices, Device characterization and geometrical details.

**Figure S2** depicts the transport characteristics of devices C, D and E studied in this work.

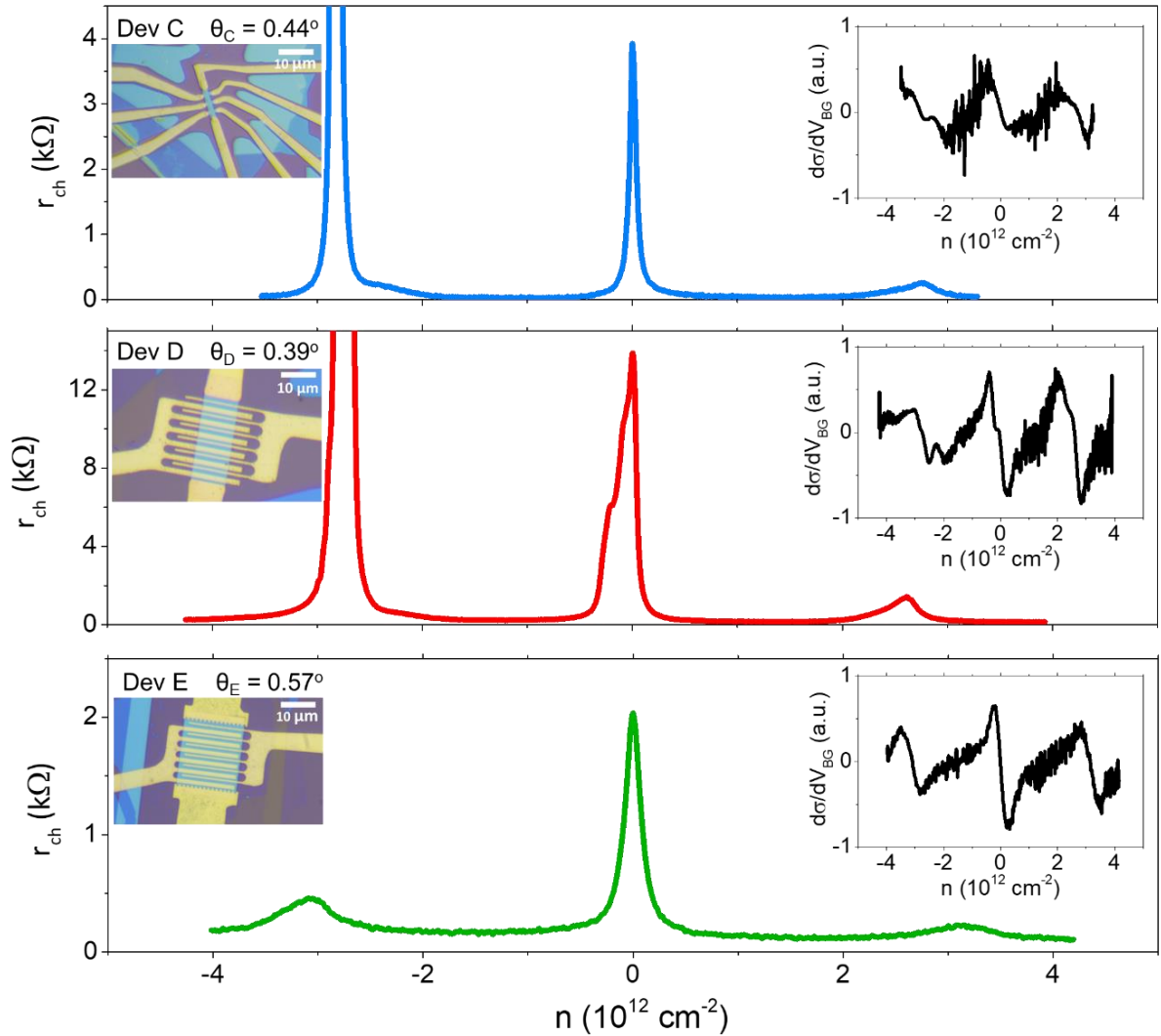

**Figure S2: Electrical characteristics of additional graphene moiré devices C-E.** Measured channel resistance, $r_{ch}$, as a function of the carrier density, n, for the three additional devices of the study at 10K. From top to bottom panels: device C (multiple-cross MLG/hBN photodetector with a θ = 0.44°), device D (interdigitated dual-grating gate MLG/hBN device with a θ = 0.39°), device E (interdigitated dual-grating gate BLG/hBN device with a θ =0.57°). Left insets in each panel show the corresponding optical image of the moiré THz device. Right insets in each panel show the variation of the channel conductivity $\sigma$ w.r.t. the gate voltage, $d\sigma/dV_G$ as a function of the normalized back-gate voltage.

In addition, detailed information about the geometrical parameters of all three device architectures fabricated for this study are highlighted in **Figure S3**.

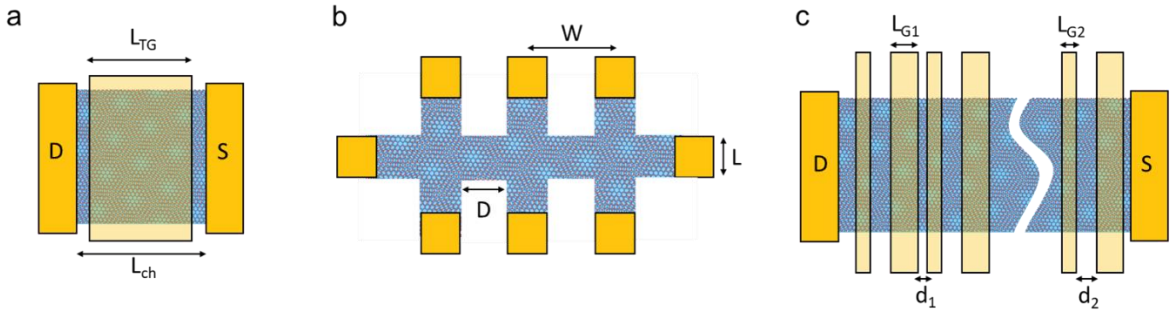

**Figure S3. Schematic views of the fabricated moiré graphene-based superlattices devices. a**, Schematic top view of the Short-Channel (SC) architecture (devices A and B). Dimensions are $L_{ch}$ = 6 (3) µm and $L_{TG}$ = 4.8 (2.5) µm for the device A (B). **b**, Schematic top view of the multicross (MC) bar geometry (device C). Dimensions are W = 2.5 µm and D = L = 1 µm. **c**, Schematic top view of the Ratchet type architecture of the Devices D and E. Top gates are characterized by an unit cell with a period of L = 3.75 µm ($L_{G1}$ = 1.5 µm, $L_{G2}$ = 0.75 µm, $d_1$ = 0.5 µm, $d_2$ = 1 µm). The unit cell was repeated 6 times.

# Note 3 – Mobility of the measured devices.

The resistance of the device channel $r_{ch}$, the contact resistance $r_c$ and mobility values $\mu$ for each device have been calculated from the measured resistance, using the procedure described by Ref. 2 and taking into account that the carrier density value $n$ is given by the expression $n = C_c(V_{BG} - V_{DP})/e$, where $C_c$ is the capacitance of the device, $e$ is the electron charge, $V_{BG}$ is the applied back-gate potential and $V_{DP}$ is the gate-potential at which the resistance maxima is obtained (indicating the position of the main Dirac point). In device C, $r_c$ was set to zero as its electrical characteristics were measured in a four-terminal configuration. The dc channel conductivity is extracted using the formula $\sigma = (L/W)/(r_{ch} - r_c)$, where $L$ is the channel length, W is the channel width.

**Figure S4** shows the obtained carrier mobility values, $\mu(n)$, for all our devices at T=10K. We consistently observe carrier mobilities reaching values higher than 10 m$^2$/Vs for holes and electrons in all our samples (see further details in **Table S2**).

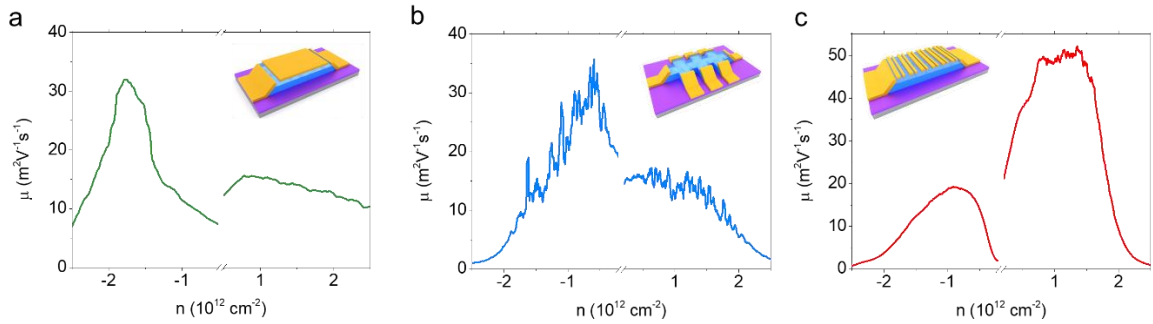

**Figure S4. Carrier mobility in our devices at T=10K.** Carrier mobility as a function of the carrier density for **a,** short channel device (device A), **b,** multi-cross device (device C) and **c,** interdigitated dual-grating gate device (device D). Insets in the panel show the corresponding architecture of the devices.

|          | $V_{DP}$ (V) | $r_c$ (Ω) | $\mu_{e,max}$ (m$^2$/Vs) | $\mu_{h,max}$ (m$^2$/Vs) |
|----------|--------------|-----------|--------------------------|--------------------------|
| Device A | 0.9          | 380       | 15.7                     | 32.1                     |
| Device B | 0.3          | 580       | 11.17                    | 12.91                    |
| Device C | 1.8          | -         | 16.3                     | 35                       |
| Device D | 2.4          | 125       | 51                       | 19.8                     |
| Device E | -1.3         | 67        | 24                       | 18                       |

**Table S2.** Electrical parameters at 10K. Gate-voltage at which the resistance maxima occur in the sample $V_{DP}$, contact resistance $r_c$, and maximum carrier mobility measured in the device at different densities for electrons $\mu_{e,max}$ and holes $\mu_{h,max}$ charge carriers.

# Note 4 – Extraction of alignment angle between graphene and hBN in the devices used for this work.

**Table S3** summarizes the moiré wavelength, $\lambda_M$, and twist angle, $\theta$, extracted in our devices from transport and Raman spectroscopy measurements. Values obtained from both techniques are consistent, with the largest disagreement between average twist angles extracted from both type of measurements being 0.06 degrees.

|          | Raman spectroscopy | | Transport measurements | |
|----------|---------------------|---------------------|---------------------|---------------------|
|          | $\lambda_M$ (nm) | $\theta$ (degrees) | $\lambda_M$ (nm) | $\theta$ (degrees) |
| Device A | 7.6   | 1.56 | 7.4   | 1.62 |
| Device B | 10.27 | 0.93 | 10.4  | 0.9  |
| Device C | 12.84 | 0.42 | 12.75 | 0.44 |
| Device D | 12.96 | 0.4  | 12.98 | 0.39 |
| Device E | -     | -    | 12.15 | 0.57 |

**Table S3.** Extracted alignment parameters from Raman spectroscopy and transport measurements

### A. *via Raman measurements*

Raman spectroscopy measurements were carried out at room temperature in air using a 532 nm laser with an incident power of 1 mW. The relative rotation angle $\theta$ and the moiré wavelength of graphene superlattices $\lambda_M$ made of monolayer graphene shown in **Table S3** are extracted from the full width at half maximum ($FWHM$) of the measured Raman 2D peak, which is given by[3]

$$FWHM \approx 5 + 2.6\lambda_M \quad \textbf{Eq. S1}$$

Moreover, the relation between $\lambda_M$ and $\theta$ is:

$$\lambda_M = \frac{(1+\xi)a}{\sqrt{2(1+\xi)(1-cos\theta)+\xi^2}} \quad \textbf{Eq. S2}$$

where $a$ is the graphene lattice constant and $\xi$ is the lattice mismatch between the graphene and the hBN lattices. **Figure S5** shows the measured 2D peak for devices A-D as well as the gaussian fits to these data used to extract the FWHM values. Specifically, we have obtained FWHM values of 24.77 cm$^{-1}$ for device A, 31.7 cm$^{-1}$ for device B, 38.4 cm$^{-1}$ for device C and 38.7 cm$^{-1}$ for device D. We further note that samples with relatively uniform twist angle are selected for this study. To do so, we measure the FWHM of the 2D peak[3,4] in five different positions per sample, verifying that such values are the same within the experimental error of our Raman spectrometer (±2cm$^{-1}$).

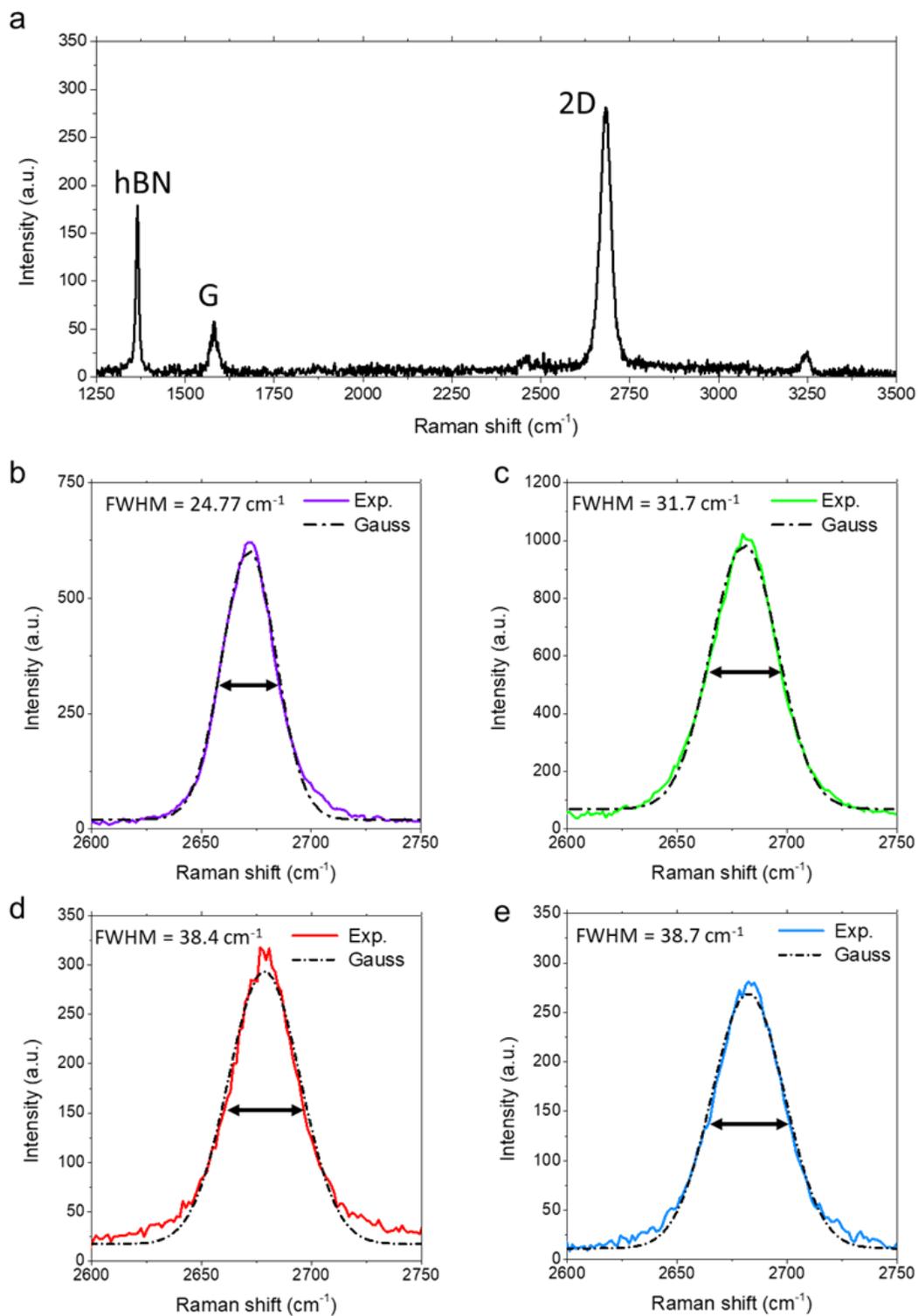

**Figure S5:** Raman spectra of the moiré graphene heterostructures. **a,** Typical Raman spectra of an aligned hBN/graphene superlattice. **b-e,** Gauss fitting of the 2D peak for the devices A, B, C and D respectively highlighting the value of the full width at half-maximum (*FWHM*).

## B. *via transport measurements*

One can obtain $\lambda_M$ and $\theta$ via transport measurements, taking into account[5] Eq. S2 and the fact that the resistance peaks associated to the satellite Dirac points occur at a charge density $n_{sp} = 4/A$, where $A = \sqrt{3}\lambda_M^2/2$ is the size of the moiré unit cell and the pre-factor 4 considers the spin and valley degeneracies in graphene.

From the gate potential $|V_{BG}-V_{DP}|$ at which the satellite points occur, $n_{sp}$ in our devices is estimated to be $n_{sp}$ is $8.4 \cdot 10^{12}$ cm$^{-2}$ for device A, $4.27 \cdot 10^{12}$ cm$^{-2}$ for device B, $2.84 \cdot 10^{12}$ cm$^{-2}$ for device C, $2.74 \cdot 10^{12}$ cm$^{-2}$ for device D and $3.3 \cdot 10^{12}$ cm$^{-2}$ for device E.

# Note 5 – THz photoresponse in our devices: additional data and analysis

*Additional responsivity data*

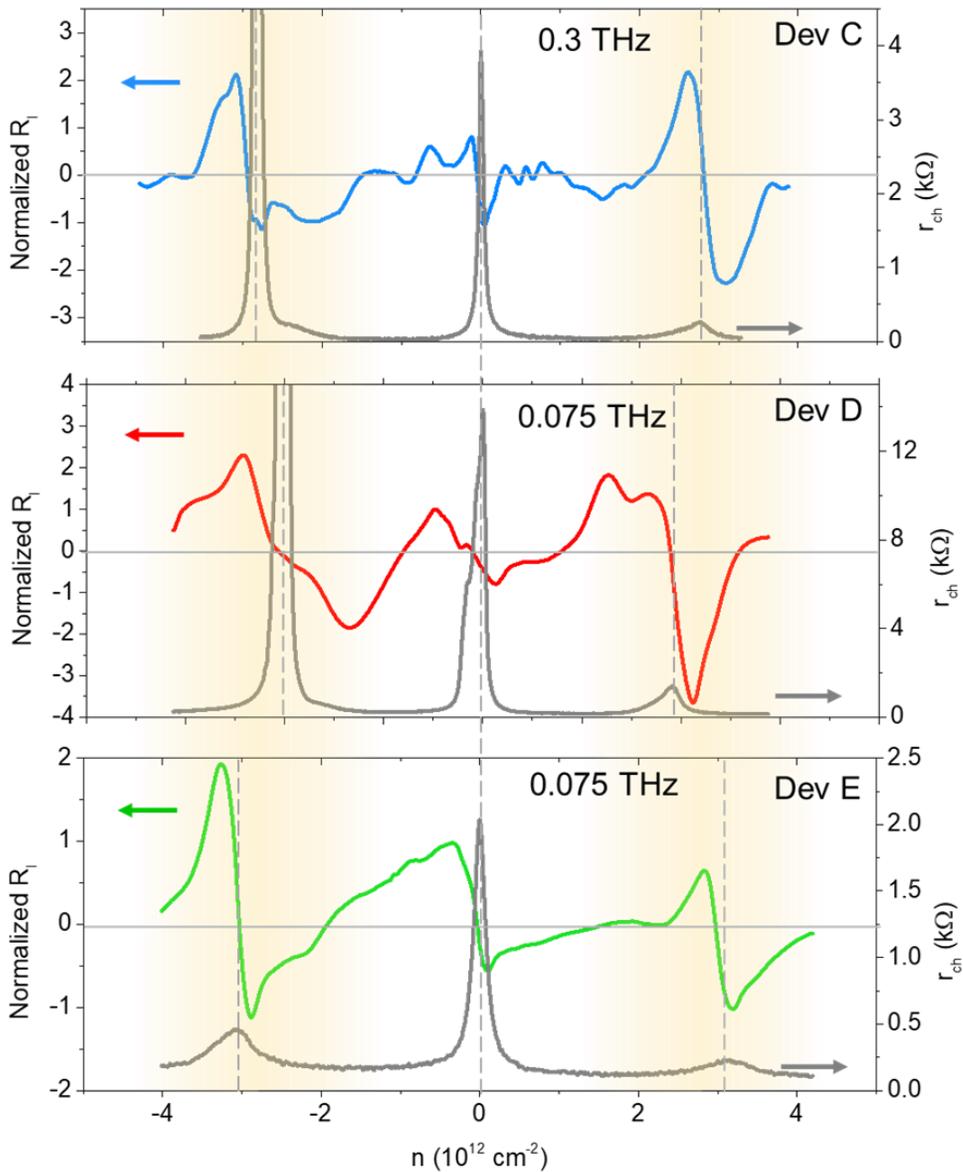

**Figure S6.** Low temperature photocurrent measurements. (Left axis) Normalized Photocurrent responsivity, $R_I$ and (right axis) measured channel resistance, $r_{ch}$, as a function of the carrier density, n, for the Devices C (hBN/MLG/hBN), D (hBN/MLG/hBN) and E (hBN/BLG/hBN) under excitation of THz radiation. The temperature of the measurement is T = 10 K in all the panels.

*Responsivity ratios for all our devices at 10K*

Table S4 presents the comparison between the THz responsivities around the two sDPs ($\delta R_I^{sDP,h}$ or $\delta R_I^{sDP,e}$) with respect to the responsivity close to the main DP ($\delta R_I^{DP}$), in all moiré graphene/hBN photodetectors measured in this study. Such ratios are shown for two excitation frequencies (0.15 THz and 0.3 THz). The data show consistently responsivity ratios above unity in devices with twist angles smaller than 1 degree.

|  | 0.15 THz | | 0.3 THz | |
|---|---|---|---|---|
| Device (twist angle) | $\delta R_I^{sDP,h}/\delta R_I^{DP}$ | $\delta R_I^{sDP,e}/\delta R_I^{DP}$ | $\delta R_I^{sDP,h}/\delta R_I^{DP}$ | $\delta R_I^{sDP,e}/\delta R_I^{DP}$ |
| A (1.62°) | 1.06 | 0.72 | 0.63 | 0.2 |
| B (0.9°) | 1.75 | 3.4 | 1.26 | 2* |
| C (0.44°) | - | - | 1.76 | 2.40 |
| D (0.39°) | 1.76 | 3.48 | 2.28 | 2.20* |
| E (0.57°) | 1.83 | 1.29* | 4.91 | 2.91* |

**Table S4**. Enhancement factor for all studied devices at 10 K. Note that, whereas a photocurrent of an intraband origin always occurs at the valence band sDP, an intraband photocurrent may take place in the electron band sDP (cases marked with an asterisk ´*´ in the table).

*Evolution of $R_I(n)$ with temperature*

Figure S7 shows the evolution of the responsivity ratio (or enhancement factor) with temperature for our aligned graphene/hBN devices (twist angles θ<1º, devices B, C, D and E) at 0.3 THz. Here, we focus our attention on the ratio between the valence band sDP and the main DP ($\delta R_I^{sDP,h}/\delta R_I^{DP}$), where the measured photocurrent is exclusively of an intraband origin.

Overall, $\delta R_I^{sDP,h}/\delta R_I^{DP}$ decreases when increasing the temperature and shows enhancement factors >1 for temperatures *T* below 80-120K in all moiré devices, irrespectively if they are made of monolayer (Figure S7a) or bilayer (Figure S7b) graphene. This is a general behavior which can be understood by the presence of electron–electron Umklapp scattering as dominant scattering mechanism close to the sDPs in graphene-based moiré superlattices at high temperatures[6].

As an additional comment, we also observe that $\delta R_I^{sDP,h}/\delta R_I^{DP}$ remains > 1 up to 120K for the moiré device made from bilayer graphene (all studied monolayer graphene devices present ratios $\delta R_I^{sDP,h}/\delta R_I^{DP}$ >1 for temperatures <100K). Such trend may be related to the fact that, due to the presence of additional intrinsic scattering sources including shear phonon scattering, the responsivity $\delta R_I^{DP}$ decreases more rapidly in samples made of bilayer graphene than in those made of monolayer graphene when increasing the temperature[7].

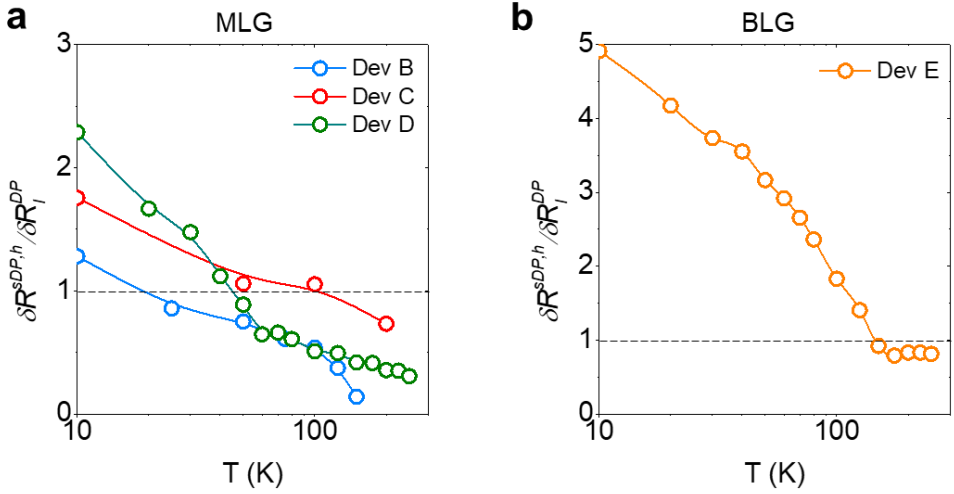

**Figure S7: Evolution of the responsivity ratio $\delta R_I^{sDP,h}/\delta R_I^{DP}$ in our aligned devices (θ<1°) at 0.3 THz.** $\delta R_I^{sDP,h}/\delta R_I^{DP}$ is plotted as a function of temperature for aligned (a) monolayer devices B, C, D and (b) bilayer device E. In all graphs, experimental data points are connected by lines to guide the eye.

# Note 6 – Estimation the enhanced responsivity ratio $\delta R_I^{sDP}/\delta R_I^{DP}$.

The broadband photoresponse of photodevices governed by intraband transitions follows the expression[8-11]:

$$R_I = -\frac{\delta U^2}{4P}\frac{d\sigma}{dV_{BG}} \qquad \textbf{Eq. S3}$$

where $\delta U$ is the amplitude of the ac potential induced by the THz radiation in the device channel, $P$ is the incident radiation power and $\sigma$ is the dc conductivity of the material.

In order to estimate the ratio $\delta R_I^{sDP}/\delta R_I^{DP}$, we assess the ratios of the two main terms of the photocurrent responsivity in Eq.S3: $\delta U$ and $d\sigma/dV_{BG}$.

Estimation of term $d\sigma/dV_{BG}$
First, we note the equivalence between $d\sigma/dV_{BG}$ and $d\sigma/dn$ in our devices is given by $d\sigma/dV_{BG} = \frac{C_c}{e}d\sigma/dn$, since the relation $n(V_{BG})$ in devices with a notable dielectric thickness > 20 nm is primarily dictated by the classical capacitance $C_c$[12] (i.e. the quantum capacitance contribution is not relevant in our devices).
In the diffusive transport regime, the dc conductivity in monolayer graphene is given by the Einstein relation[13]:

$$\sigma = e^2 D(E_F) v_F^2 \tau/2 \qquad \textbf{Eq. S4}$$

where $D(E_F) = (\sqrt{g_s g_v n})/(\sqrt{\pi}\hbar v_F)$ is the density of states at the Fermi level $E_F$, $v_F$ is the Fermi velocity of charge carriers, $\tau$ is the carrier density dependent scattering time, and $g_s$, $g_v$ are the spin and valley degeneracies of graphene, respectively.

In a diffusive system, the two main scattering events in the system can have a long- or short-range origin, and depend on $n$ as $\tau \propto \sqrt{n}$ and $\tau \propto 1/\sqrt{n}$, respectively. These two mechanisms give rise to two contributions to the total conductivity $\sigma$ (see Eq.S5), which are linear or independent on the carrier density $n$ for long ($\sigma_l$) and short-range ($\sigma_s$) scattering events, respectively.

$$\frac{1}{\sigma(n)} = \frac{1}{\sigma_l(n)} + \frac{1}{\sigma_s} \qquad \textbf{Eq. S5}$$

We remark that the two conductivity contributions $\sigma_l$, $\sigma_s$ encompass different scattering mechanisms that may occur in any graphene device[14-16].

In the case of graphene/hBN moiré superlattice devices, their total conductivity $\sigma^{SL}(n)$ can be estimated by considering three contributions $\sigma$, one from the main DP and two form the satellite DPs as[17]:

$$\frac{1}{\sigma^{SL}(n)} = \frac{1}{\sigma(n)} + \frac{1}{\sigma(n+n_{sDP})} + \frac{1}{\sigma(n-n_{sDP})} \qquad \textbf{Eq. S6}$$

**Figure S8** shows the normalized estimated conductivity $\sigma^{SL}(n)$ and its variation with respect to the carrier density of the channel $d\sigma^{SL}/dn$.

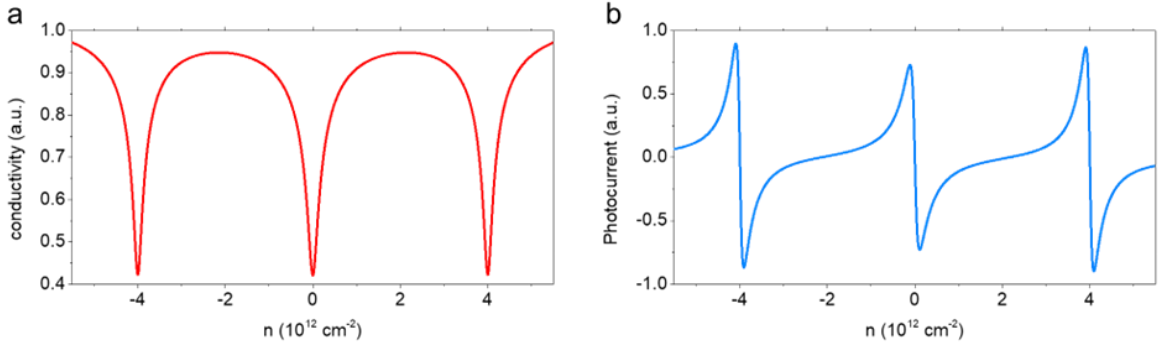

**Figure S8. a,** Estimated dependence of the conductivity with $n$ in graphene moiré superlattice devices $\sigma^{SL}$ calculated from Eq.S6 and **b,** photocurrent calculation via $d\sigma^{SL}/dn$.

Importantly, $d\sigma^{SL}/dn$ shows a similar magnitude of the photoresponse close to the main and satellite Dirac points, which would result in a responsivity ratio $\delta R_I^{SP}/\delta R_I^{DP}$ close to unity. More explicitly, the estimated responsivity ratio from this factor is roughly given by:

$$\left(\frac{d\sigma}{dn}\right)^{(sDP)} / \left(\frac{d\sigma}{dn}\right)^{(DP)} \approx \left(\sqrt{g_v^{(sDP)}}\, v_F^{(sDP)}\right) / \left(\sqrt{g_v^{(DP)}}\, v_F^{(DP)}\right) \qquad \textbf{Eq. S7}$$

In the former expression, $v_F^{(DP)}, v_F^{(sDP)}$ and $g_v^{(DP)}, g_v^{(sDP)}$ are the Fermi velocities and valley degeneracies of graphene carriers close to the main and satellite Dirac points, respectively. The maximum ratio $\delta R_I^{SP}/\delta R_I^{DP}$ that can be estimated from Eq. S7 is ~1.26 for the case $v_F^{(sDP)} \sim 0.73\, v_F^{(DP)}$ and $g_v^{(sDP)} = 3 g_v^{(DP)}$ (maximum values reported in literature for these electronic parameters[18,19]). Such ratio is, therefore, close to 1 and notably smaller than the ones observed experimentally in our devices with θ <1°. As such, the term $d\sigma/dV_{BG}$ alone cannot explain the observed enhancement of the measured photocurrent near the satellite Dirac points.

*AC potential induced in the channel by the THz radiation, $\delta U$*

The term $\delta U$ is commonly approximated in literature[8-11] by the constant value $U_a$, amplitude of the ac potential between the lobes of the antenna. Nonetheless, in practical devices, $\delta U$ depends on the coupling channel material. In this sense, we note that $\delta U$ can be related to the carrier density oscillation in the material $\delta n$ via[20] $\delta U = (e^2/C)\delta n$. In the former expression, $C$ is the dynamic capacitance and $\delta n$ can be approximated to the Fermi level $E_F$ and the equilibrium carrier density of the channel $n_{eq}$ as[21] $\delta n = (\partial n_{eq}/\partial E_F)\partial E_F$. Moreover, to a first approximation (zero temperature and zero carrier density), the compressibility $\partial n_{eq}/\partial E_F$ is given by the density of states at the Fermi level of the system $D(E_F)$.

In this sense, $\delta n$ is different at Fermi levels close to the main and satellite Dirac points in a graphene moiré superlattice device. In consequence, the ratio $\delta R_I^{sDP}/\delta R_I^{DP}$ can be estimated as:

$$\frac{\delta U^{2(sDP)}}{\delta U^{2(DP)}} = \frac{\left(D(E_F)^{(sDP)}\right)^2}{\left(D(E_F)^{(DP)}\right)^2} = \frac{g_v^{(sDP)}\left(v_F^{(DP)}\right)^2}{g_v^{(DP)}\left(v_F^{(sDP)}\right)^2} \qquad \textbf{Eq. S8}$$

By inserting Eq. S7 and S8 in Eq. S3, we have

$$\frac{\delta R_I^{sDP}}{\delta R_I^{DP}} \approx \frac{\left(g_v^{(sDP)}\right)^{3/2} v_F^{(DP)}}{\left(g_v^{(DP)}\right)^{3/2} v_F^{(sDP)}} \qquad \textbf{Eq. S9}$$

which is Eq.1 in the main text.

# Note 7 – Assessment of energy gaps at the main and satellite DPs via transport measurements

In this note, we assess the size of the energy gaps at the main and satellite Dirac points in one of our devices via temperature dependent transport measurements. This is a conventional technique used in literature to extract the energy gap of graphene moiré superlattice systems (see e.g. Ref 4). **Figure S9a** shows the evolution of $r_{ch}(n)$ at three different temperatures from 10K to 250K for device D ($\theta \sim 0.4$ degrees). The resistance maxima $r_{ch}^{max}$ gets larger at lower temperatures at the hole band SDP and the main DP. Arrhenius plots showing the dependence of the minimum conductance of the device ($1/r_{ch}^{max}$) with $1/T$ at the hole band SDP and main DP are depicted in **Figures S9b** and **S9c**, respectively.

From these data, the extracted energy gaps at these two DPs are $\Delta_h \sim 21$ meV and $\Delta \sim 23$ meV, respectively, values which are in good agreement with those probed in literature in similar samples via transport measurements[5]. More importantly, the values of $\Delta_h$ extracted via the here proposed THz photocurrent spectroscopy (**Figure 3** in the main text) are in excellent agreement with those extracted via transport measurements (**Figure S9b**). In contrast, $r_{ch}^{max}$ is approximately constant at the conduction band SDP for the measured temperature range (it varies between 0.6 to 0.7 k$\Omega$), behaviour which is also observed in additional graphene moiré superlattices reported in literature[22]. The fact that $r_{ch}^{max}$ does not increase monotonically when lowering the temperature as expected for thermally activated transport is a direct consequence of the lack of an overall bandgap present around the satellite Dirac point in the conduction band (see calculated bandstructure in **Figure 5b** of the main text).

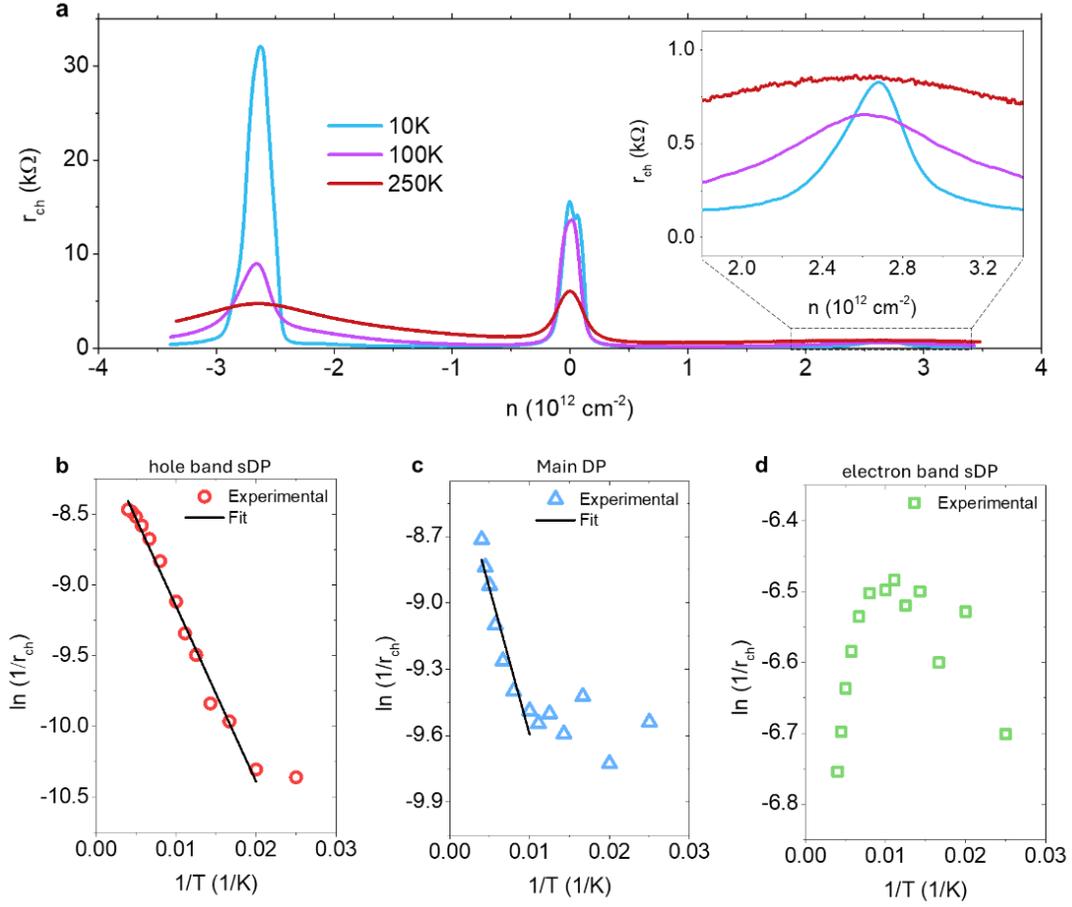

**Figure S9: Evolution of channel resistance r$_{ch}$ with temperature for device D. a**, r$_{ch}$(n) at three selected temperatures. **b,** Value of the minimum conductance (1/r$_{ch}^{max}$, with r$_{ch}^{max}$ being the resistance maximum) measured in the channel close to the hole band sDP. The black line corresponds to a fit of the minimum conductance assuming thermal activated transport 1/r$_{ch}^{max}$ ∝ exp(-Δ$_h$/2K$_B$T), with Δ$_h$ = 21 meV. **c,** Value of the minimum conductance (1/r$_{ch}^{max}$, with r$_{ch}^{max}$) measured in the channel close to the main DP. The black line corresponds to a fit of the minimum conductivity assuming thermal activated transport 1/r$_{ch}^{max}$ ∝ exp(-Δ/2K$_B$T), with Δ = 23 meV. **d,** Value of the minimum conductance (1/r$_{ch}^{max}$, with r$_{ch}^{max}$ being the resistance maximum) measured in the channel close to the conduction band sDP.

# Note 8 – Assessment of energy gaps at the electron-band SDP via THz photocurrent spectroscopy.

**Figure S10** shows the responsivity $R_I(n)$ of different devices measured close to the electron-band sDP at different sub-THz frequencies ($f$ = 0.075, 0.15, 0.3 and 0.6, respectively). Here, we can observe how $R_I(n)$ shows an intra-band photocurrent at all measured frequencies in devices with a large rotation angle (θ>1˚) between the graphene and hBN crystals (device A, see **Figure S10a**).

In contrast, devices with a lower rotation angle (θ<1˚) behave in a different way. Whereas $R_I(n)$ of devices D and E (see **Figure S10 b** and **c**) also show an intraband type of photoresponse at the lowest frequency (0.075THz); the lineshape of $R_I(n)$ evolves towards an interband-type of photorresponse at larger frequencies, above 0.3 THz and 0.15THz for devices D and E respectively.

From these measurements, we extract an approximate size of the energy gap to be $\Delta_e \approx$1.2 meV and $\Delta_e \approx$0.6 meV for devices D (θ~0.39˚) and E (θ~0.57˚), respectively.

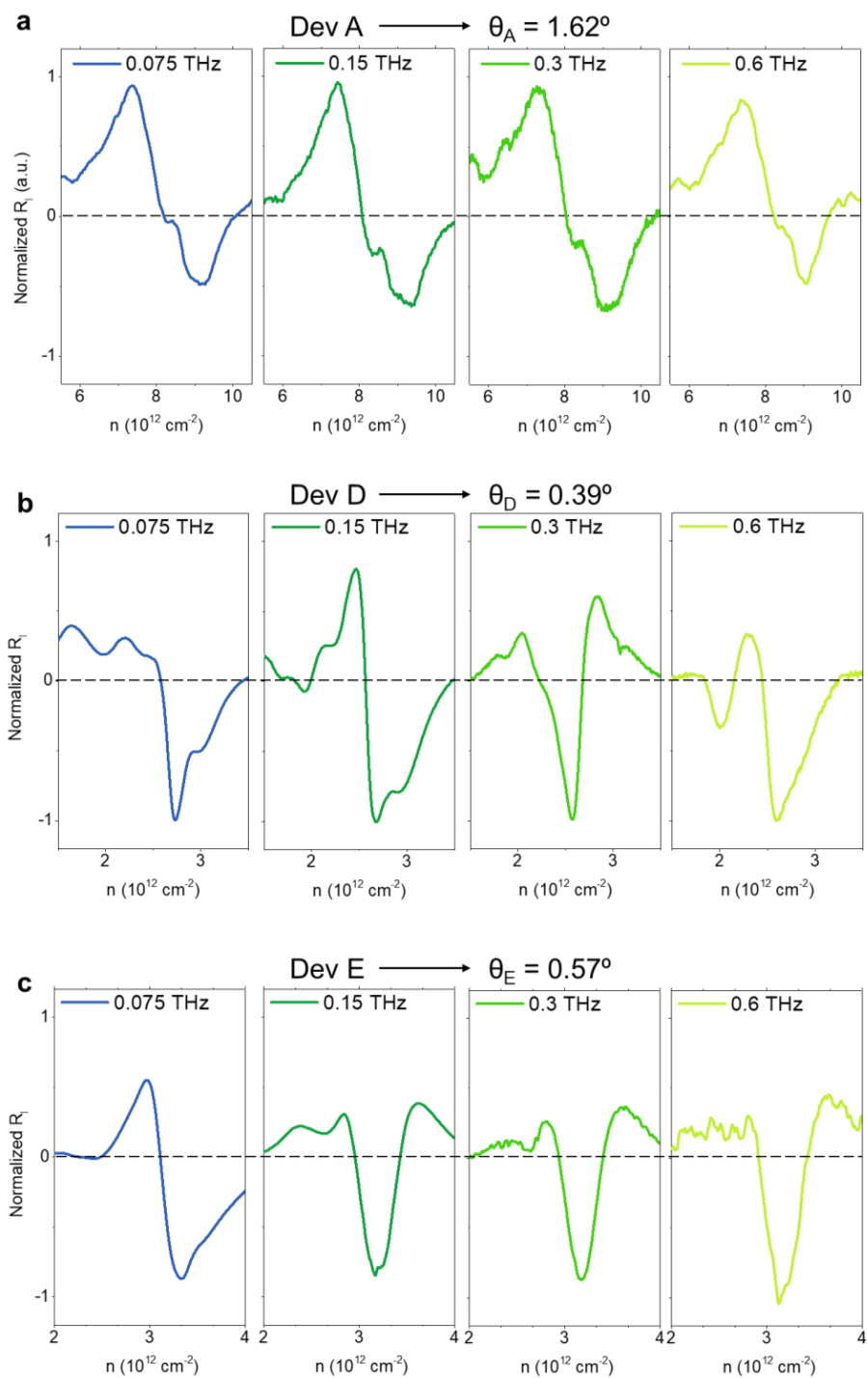

**Figure S10. Low temperature photocurrent spectroscopy at the electron-band SDP.** Normalized photocurrent responsivity $R_I$ at sub-THz frequencies as a function of the carrier density, n, close to the electron-band SDP for **a**, device A, **b**, device D and **c**, device E. All measurements are undertaken at 10K

# Note 9 – Additional calculations.

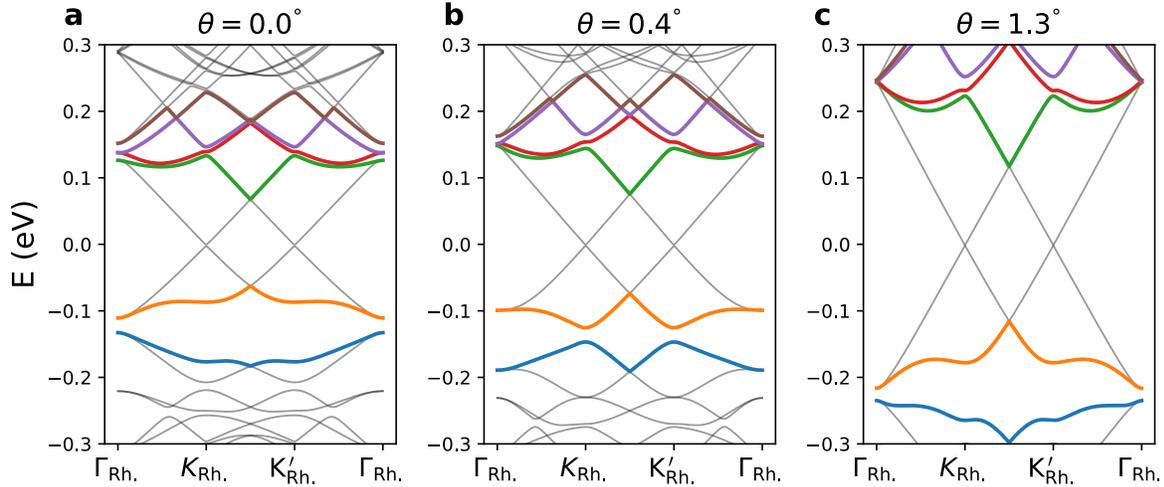

**Figure S11. Band structures of graphene/hBN heterostructures with different twist angles.** Electronic bands in a wider energy range around half-filling, showing the principal and secondary Dirac points in each case. The thicker, colored bands correspond to those relevant to transitions near the sDPs which are discussed in the main text. The three cases shown are **a,** untwisted, **b,** the case of $\theta = 0.44°$ also shown in **Figures 5a** and **5b** of the main text, and **c**, a larger twist angle of $\theta = 1.3°$.

**Figure S11** shows the electronic bands near half-filling of three graphene/hBN systems with different twist angles. Panel **a** shows the untwisted case, in perfect agreement with similar calculations from Ref. 23. Panel **b** reproduces the bands for the $\theta = 0.44°$ case discussed in the main text, but now over a wider range of energies than those shown in the sDP zooms of **Figure 5a** and **Figure 5b**. Panel **c** shows the bands for the largest twist angle ($\theta = 1.3°$) for which energy gaps were extracted for Fig 4c. We note that the sDPs shift to higher energies as the twist angle increases, as noted previously[23]. We also note that the folding of graphene's Brillouin Zone (BZ) into the smaller BZ of the rhombohedral (Rh.) supercell used for tight-binding calculations can lead to an apparent change in the shape of the bands near the sDPs. This is particularly evident in the shapes of the blue and orange bands (labelled α and β in the main text) near the hole sDP. Although the K and K' points of graphene are folded to K and K' points of the rhombohedral BZ, they can swap places depending on the exact unit cell size. This leads to, for example, the smaller gap at the valence sDP ($\Delta_h^1$), moving between $\Gamma_{Rh.}$ (panels **a** and **c**) and $K_{Rh.}$ (panel b) in these band structures.

Finally, we observe that only minute band gaps are found at the main DP in our model. The larger gap observed here in experiments require additional terms in the Hamiltonian to capture effects such as interaction and relaxation[24]. As these terms play a less prominent role at the sDPs, which are the main focus of this work, we have not included them in our calculations.

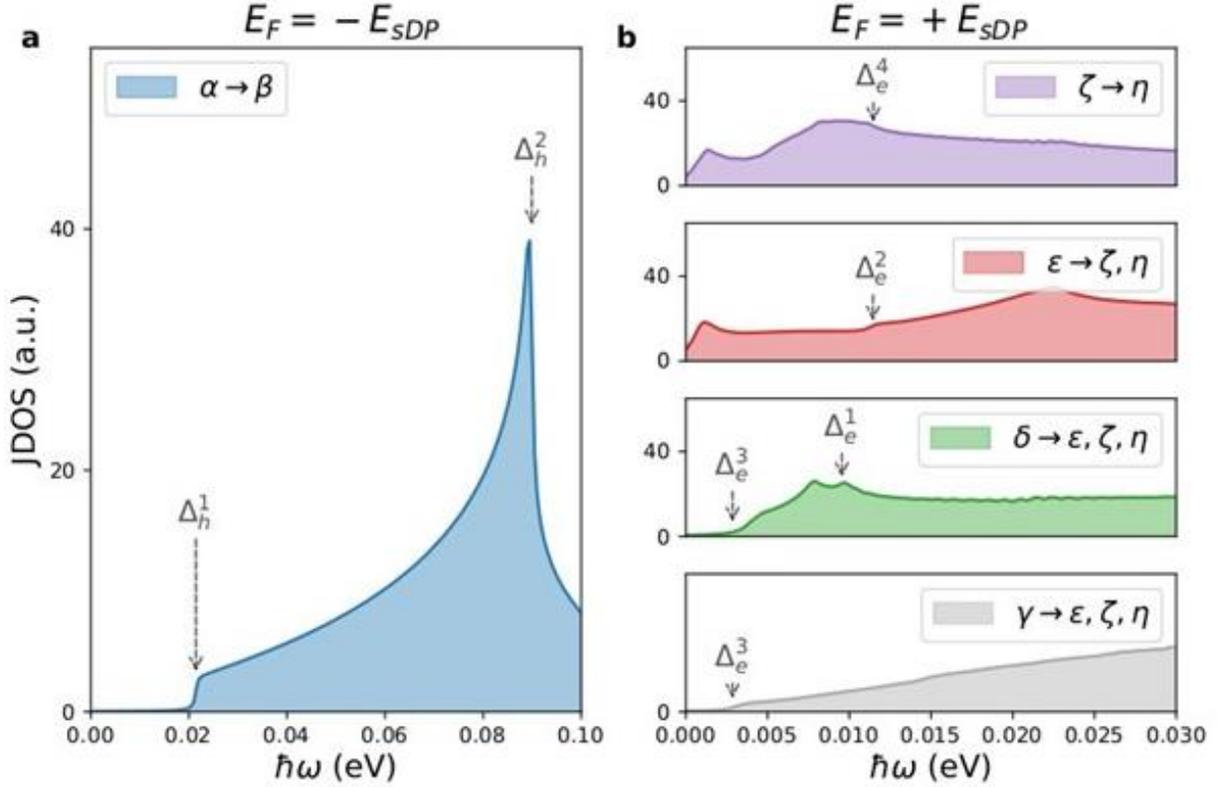

**Figure S12. JDOS and interband conductivity of the system. a,** JDOS between bands $\alpha$ and $\beta$ (**Figure 5a,** main text), calculated as a function of the excitation energy $\hbar\omega$, corresponding to transitions near $E_F = -E_{sDP}$ (valence band sDP) for a device with θ=0.4°. **b,** JDOS calculated between the different bands in **Figure 5b,** main text as a function of the excitation energy $\hbar\omega$, corresponding to transitions near $E_F = +E_{sDP}$. The legends show the bands included in the calculations.

**Figure S12** shows the Joint Density of States (JDOS) corresponding to the possible band transitions and excitation energy ranges relevant to the **a**, valence and **b**, conduction sDP photoresponses discussed in **Figure 5** of the main text. The JDOS associated with transitions between a pair of bands *m* and *n*, with excitation energy $\hbar\omega$, is given by

$$\text{JDOS}_{(m \to n)}(\hbar\omega) = \int d\boldsymbol{k}\, \delta(\varepsilon_n(\boldsymbol{k}) - \varepsilon_m(\boldsymbol{k}) - \hbar\omega) \qquad \textbf{Eq. S10}$$

where $\varepsilon_n(\boldsymbol{k})$ is the energy of band *n* at a point $\boldsymbol{k}$ in reciprocal space. The legends in panels **a** and **b** indicate the bands considered in each panel, where we note that, due to the larger number of bands available, transitions to multiple bands are grouped together in panel **b**.

For Fermi energies near the valence sDP, the only relevant transitions are between the band directly below (α) and above (β) the gap. The onset of a finite JDOS in **Figure S12a** coincides with the minimum gap value between these bands ($\Delta_h^1$), with a large peak observed at $\hbar\omega = \Delta_h^1$: this can be associated with transitions near $\Gamma_{Rh.}$ where both bands are reasonably flat. Both the onset and peak in JDOS can also be clearly observed in the optical conductivity in panel **c**.

The situation is more complicated near the conduction side sDP, due to the larger number of bands which can play a role in transitions, including some (e.g. ε, ζ) which can play the role of either the valence (source) or conduction (destination) band, depending on the exact Fermi energy and ***k*** point. For clarity, in **Figure S12b** we group (and color) transitions in subpanels by their valence band. While there are features at excitation energies corresponding to the important energy gaps ($\Delta_e^1, \Delta_e^2, \Delta_e^3, \Delta_e^4$), they are less prominent than those near the valence side sDP, and are more difficult to directly connect to the optical conductivity (blue curve in panel **d.**). This can be understood by considering the following points:

- The optical conductivity is composed of all possible transitions at a particular excitation energy. The possible transitions overlap considerably in their energy ranges of possible excitation energies, so are harder to disentangle.
- Unlike the optical conductivity, which is for a fixed Fermi energy, JDOS captures all possible transitions between two bands. This is why the ζ → η JDOS in the top panel of **Figure S12b** has finite contributions for $\hbar\omega < \Delta_e^4$, which corresponds to the gap between them at the conduction sDP. These two bands approach each other and eventually touch at higher energies, giving finite JDOS for smaller excitation energies which are irrelevant at $E_F \approx +E_{sDP}$.

The contribution of individual pairs of bands to the shift conductivity is discussed in the main text.

## Note 10– Noise equivalent power.

We have evaluated the Noise-Equivalent-Power (NEP) in three of our detectors with θ <1˚ under excitation of THz at 10K. Since we have performed zero-bias photocurrent experiments, the Johnson–Nyquist noise, known as thermal noise, is the principal source of noise that we considered for the calculations. Thus, using Johnson–Nyquist relation for the noise spectral density, $N = \sqrt{4k_B T r_{ch}}$, the NEP can be calculated using the formula[8,10]:

$$NEP = \frac{\sqrt{4k_B T r_{ch}}}{R_I r_{ch}} \qquad \textbf{Eq. S11}$$

Interestingly, we observe an enhancement of the performance (i.e. lower values of NEP) in the vicinity of the SDPs in the examined moiré superlattices detectors with θ <1˚ (see **Figure S13**). We have quantified such reduction of the NEP by taking into account the minimum observable value of the NEP close to the main Dirac point ($NEP_{min,DP}$) and the minimum value of the NEP close to the satellite Dirac points ($NEP_{min,sDP}$). We have estimated an enhancement of the performance with ratios, $\delta NEP = NEP_{min,DP}/NEP_{min,sDP}$, down to ~0.2. This reduction of the NEP demonstrates that graphene moiré superlattices devices can be used as sensitive and low noise THz detectors.

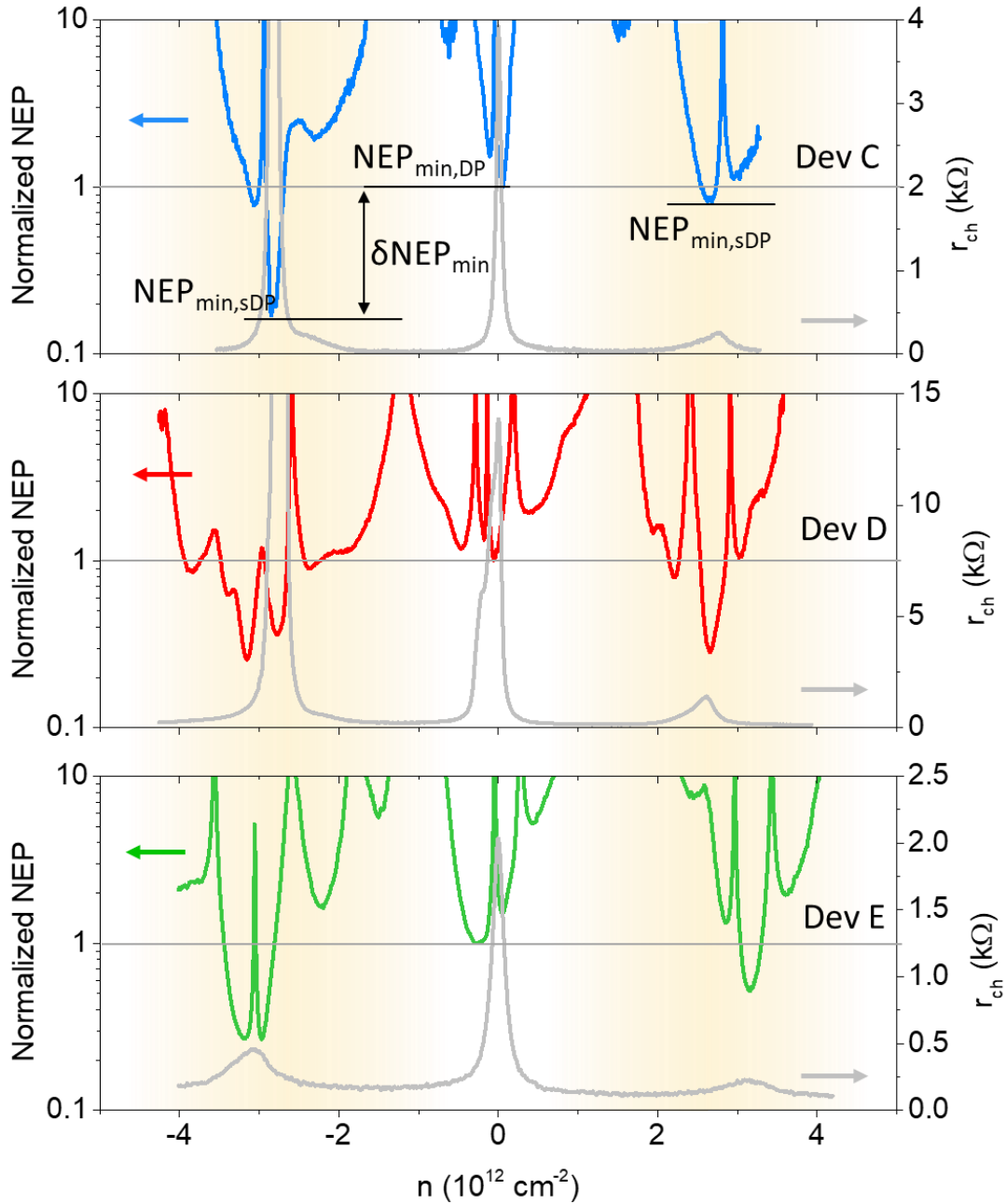

**Figure S13. Low temperature Noise-equivalent-power of the moiré superlattices.** Calculated NEP (Left) as a function of the carrier density, n, for the Devices C (top), D (middle) and E (bottom) at 10K under the excitation of 0.3 THz. NEP values have been normalized w.r.t. the minimum measured NEP close to the main Dirac point (horizontal grey line) for clarity. The corresponding device resistance curve (right) has been superimposed in the panels in grey color.

# SUPPLEMENTARY REFERENCES